\documentclass[11pt]{article}
\usepackage{graphicx}
\usepackage{amsmath}
\usepackage{amssymb}
\usepackage{caption2}
\begin{document}
\bibliographystyle{prsty}
\begin{center}
{\large {\bf \sc{  Analysis of  the ${3\over 2}^+$ heavy and doubly
heavy
baryon states   with QCD sum rules }}} \\[2mm]
Zhi-Gang Wang \footnote{E-mail:wangzgyiti@yahoo.com.cn.  }     \\
 Department of Physics, North China Electric Power University,
Baoding 071003, P. R. China

\end{center}

\begin{abstract}
In this article, we study the  ${3\over 2}^+$ heavy and doubly heavy
baryon states $\Xi^*_{cc}$, $\Omega^*_{cc}$,  $\Xi^*_{bb}$,
$\Omega^*_{bb}$,
 $\Sigma_c^*$, $\Xi_c^*$, $\Omega_c^*$, $\Sigma_b^*$,
$\Xi_b^*$ and $\Omega_b^*$   by subtracting the contributions from
the corresponding  ${3\over 2}^-$ heavy and doubly heavy  baryon
states with the QCD sum rules, and make reasonable predictions for
their masses.
\end{abstract}



\section{Introduction}

In 2006, the Babar collaboration reported the first observation of
the ${3\over 2}^+$ heavy baryon state $\Omega_{c}^{*}$  in
 the radiative decay $\Omega_{c}^{*}\rightarrow\Omega_c\gamma$ \cite{OmegaC}.
By now, the ${1\over 2}^+$ antitriplet states ($\Lambda_c^+$,
$\Xi_c^+,\Xi_c^0)$,  and the ${1\over 2}^+$ and ${3\over 2}^+$
sextet states ($\Omega_c,\Sigma_c,\Xi'_c$) and
($\Omega_c^*,\Sigma_c^*,\Xi^*_c$) have been well established
\cite{PDG}.

In 2008, the D0 collaboration reported the first observation of the
doubly strange baryon state $\Omega_{b}^{-}$ in the decay channel
$\Omega_b^- \to J/\psi\thinspace\Omega^-$ in $p\bar{p}$ collisions
at $\sqrt{s}=1.96$ TeV \cite{OmegabD0}. The experimental value
$M_{\Omega_b^-}=(6.165\pm 0.010\thinspace \pm 0.013)\thinspace \,
\rm{GeV}$ is about $0.1 \, \rm{GeV}$ larger than the most
theoretical calculations
\cite{Roncaglia95,HH-Valcarce,Jenkins96,Bowler96,Mathur02,Ebert05,Ebert08,Karliner07,Liu07,HH-Roberts}.
However,  the CDF collaboration did not confirm the   measured
 mass \cite{OmegabCDF}, i.e. they  observed the mass of the $\Omega^-_b$   is about $(6.0544\pm 0.0068
\pm 0.0009) \,\rm{GeV} $, which is consistent with the most
theoretical calculations. On the other hand,  the theoretical
prediction $M_{\Omega_c^0}\approx 2.7\, \rm{GeV}$
\cite{Roncaglia95,HH-Valcarce,Jenkins96,Bowler96,Mathur02,Ebert05,Ebert08,Karliner07,Liu07,Huang0811,HH-Roberts,Liu0909}
is consistent with the experimental data
$M_{\Omega_c^0}=(2.6975\pm0.0026) \,\rm{GeV}$ \cite{PDG}. The
$S$-wave bottom baryon states are far from complete, only the
$\Lambda_b$, $\Sigma_b$, $\Sigma_b^*$, $\Xi_b$, $\Omega_b$ have been
observed \cite{PDG}.

In 2002,  the SELEX collaboration reported   the first observation
of a signal for the doubly charm baryon  state  $ \Xi_{cc}^+$ in the
charged decay mode $\Xi_{cc}^+\rightarrow\Lambda_c^+K^-\pi^+$
\cite{SELEX2002}, and confirmed later by the same collaboration in
the decay mode $\Xi_{cc}^+\rightarrow pD^+K^- $ with measured mass
$M_{\Xi}=(3518.9 \pm 0.9) \,\rm{ MeV }$ \cite{SELEX2004}. However,
the Babar and Belle collaborations  have not observed any evidence
for the doubly charm baryon states in $e^{+}e^{-}$ annihilations
\cite{Xi-cc-1,Xi-cc-2}. No experimental evidences for the ${3\over
2}^+$ doubly heavy baryon states are observed \cite{PDG}.  There
have been several approaches to deal with the doubly heavy baryon
masses, such as the relativistic quark model
\cite{HH-Ebert,HH-Martynenko}, the non-relativistic quark model
\cite{HH-Roberts,HH-Albertus,HH-Vijande,HH-Gershtein}, the
three-body Faddeev method \cite{HH-Valcarce}, the potential approach
combined with the QCD sum rules \cite{HH-Kiselev}, the quark model
with AdS/QCD inspired potential \cite{HH-Giannuzzi}, the MIT bag
model \cite{HH-He}, the full QCD sum rules
\cite{HH-Narison,HH-Zhang},  the Feynman-Hellmann theorem and
semiempirical mass formulas \cite{Massform}, and the effective field
theories \cite{HH-Brambilla},  etc.

The charm  and bottom baryon states which contain one (two) heavy
quark(s) are particularly interesting for studying dynamics of the
light quarks in the presence  of the heavy quark(s), and serve as an
excellent ground for testing predictions of the
 quark models and heavy quark symmetry.
On the other hand, the QCD sum rules is a powerful theoretical tool
in studying the ground state heavy baryon states
\cite{SVZ79,PRT85,NarisonBook}.

In the QCD sum rules, the operator product expansion is used to
expand the time-ordered currents into a series of quark and gluon
condensates which parameterize the long distance properties of the
QCD vacuum. Based on the quark-hadron duality, we can obtain copious
information about the hadronic parameters at the phenomenological
side \cite{SVZ79,PRT85,NarisonBook}. There have been several works
on the masses of the heavy baryon states with the full QCD sum rules
and the QCD sum rules in the heavy quark effective theory (one can
consult Ref.\cite{Wang0912} for more literatures).

In Ref.\cite{Oka96}, Jido et al introduce a novel approach based on
the QCD sum rules to separate the contributions of   the
negative-parity light flavor  baryons from the positive-parity light
flavor baryons, as the interpolating currents may have non-vanishing
couplings to both the negative- and positive-parity baryons
\cite{Chung82}. Before the work of Jido et al, Bagan et al take the
infinite mass limit for the heavy quarks to separate the
contributions of the positive and negative parity heavy baryon
states unambiguously \cite{Bagan93}.

 In
Refs.\cite{Wang0704,Wang0809,Wang0910}, we study the
 heavy baryon states $\Omega_Q$, $\Xi'_Q$, $\Sigma_Q$, $\Omega_Q^*$, $\Xi^*_Q$ and
$\Sigma^*_Q$ with the full QCD sum rules, and observe that the pole
residues of the ${\frac{3}{2}}^+$ heavy baryons from the sum rules
with different tensor structures are consistent with each other,
while the pole residues of the ${\frac{1}{2}}^+$ heavy baryons from
the sum rules with different tensor structures differ from each
other greatly.  In Refs.\cite{Wang0912,Wang0101}, we follow
Ref.\cite{Oka96} and study the masses and pole residues of the
${\frac{1}{2}}^+$ heavy baryon states $\Omega_Q$, $\Xi'_Q$,
$\Sigma_Q$, $\Lambda_Q$ and $\Xi_Q$ by subtracting the contributions
of  the negative parity heavy baryon states to overcome  the
embarrassment. Those pole residues are important parameters in
studying the radiative decays $\Omega_Q^*\to \Omega_Q \gamma$,
$\Xi_Q^*\to \Xi'_Q \gamma$ and $\Sigma_Q^*\to \Sigma_Q \gamma$
\cite{Wang0910,Wang0909}, etc. In Ref.\cite{Wang0101-2}, we extend
our previous works  to study the ${1\over 2}^+$ doubly heavy baryon
states $\Xi_{QQ}$ and $\Omega_{QQ}$ with the full QCD sum rules.

In this article,  we study the  ${3\over 2}^+$ heavy and doubly
heavy baryon states $\Xi^*_{cc}$, $\Omega^*_{cc}$, $\Xi^*_{bb}$,
$\Omega^*_{bb}$, $\Sigma_c^*$, $\Xi_c^*$, $\Omega_c^*$,
$\Sigma_b^*$, $\Xi_b^*$ and $\Omega_b^*$   by subtracting the
contributions from the corresponding ${3\over 2}^-$ heavy and doubly
heavy baryon states with the QCD sum rules.

 The article is arranged as follows:  we derive the
QCD sum rules for the masses and the pole residues of  the heavy and
doubly heavy baryon states $\Xi^*_{cc}$, $\Omega^*_{cc}$,
$\Xi^*_{bb}$, $\Omega^*_{bb}$, $\Sigma_c^*$, $\Xi_c^*$,
$\Omega_c^*$, $\Sigma_b^*$, $\Xi_b^*$ and $\Omega_b^*$  in Sect.2;
 in Sect.3, we present the numerical results and discussions; and Sect.4 is reserved for our
conclusions.

\section{QCD sum rules for  the baryon states $\Omega^*_{QQ}$, $\Xi^*_{QQ}$,  $\Omega_Q^*$, $\Xi_Q^*$  and  $\Sigma_Q^*$ }
The ${3\over 2}^+$ heavy and doubly heavy baryon states
$\Omega^*_{QQ}$, $\Xi^*_{QQ}$, $\Omega_Q^*$,  $\Xi_Q^*$  and
$\Sigma_Q^*$ can be interpolated by the following currents
$J^{\Omega^*_{QQ}}_\mu(x)$, $J^{\Xi^*_{QQ}}_\mu(x)$,
$J^{\Omega^*_Q}_\mu(x)$, $J^{\Xi^*_Q}_\mu(x)$ and
$J^{\Sigma^*_Q}_\mu(x)$  respectively,
\begin{eqnarray}
J^{\Omega^*_{QQ}}_\mu(x)&=& \epsilon^{ijk}  Q^T_i(x)C\gamma_\mu
Q_j(x)    s_k(x) \, , \nonumber \\
J^{\Xi^*_{QQ}}_\mu(x)&=& \epsilon^{ijk}  Q^T_i(x)C\gamma_\mu Q_j(x)
 q_k(x)  \, ,  \nonumber \\
 J^{\Omega^*_Q}_\mu(x)&=& \epsilon^{ijk}  s^T_i(x)C\gamma_\mu s_j(x)  Q_k(x)  \, ,  \nonumber \\
J^{\Xi^*_Q}_\mu(x)&=& \epsilon^{ijk}  q^T_i(x)C\gamma_\mu s_j(x)  Q_k(x)  \, ,  \nonumber \\
J^{\Sigma^*_Q}_\mu(x)&=& \epsilon^{ijk}  u^T_i(x)C\gamma_\mu d_j(x)
Q_k(x) \, ,
\end{eqnarray}
where the  $Q$ represents the heavy quarks $c$ and $b$,  the $i$,
$j$ and $k$ are color indexes, and the $C$ is the charge conjunction
matrix. In the heavy quark limit, the heavy and doubly heavy baryon
states can be described by the  diquark-quark model
\cite{HH-Kiselev}.

 The corresponding ${3\over 2}^-$ heavy and doubly heavy baryon states can be
interpolated by the  currents $J^{-}_\mu =i\gamma_{5} J^{+}_{\mu}$
because multiplying $i \gamma_{5}$ to the $J^{+}_\mu$ changes the
parity of the $J^{+}_\mu$ \cite{Oka96}, where the $J^{+}_\mu$
denotes the currents $J^{\Omega^*_{QQ}}_\mu(x)$,
$J^{\Xi^*_{QQ}}_\mu(x)$, $J^{\Omega^*_Q}_\mu(x)$,
$J^{\Xi^*_Q}_\mu(x)$ and $J^{\Sigma^*_Q}_\mu(x)$.

The correlation functions $\Pi^{\pm}_{\mu\nu}(p)$ are defined by
\begin{eqnarray}
\Pi^{\pm}_{\mu\nu}(p)&=&i\int d^4x e^{ip \cdot x} \langle
0|T\left\{J^{\pm}_\mu(x)\bar{J}^{\pm}_{\nu}(0)\right\}|0\rangle \, .
\end{eqnarray}
The currents $J_{\mu}^{\pm}(x)$ couple  to both the
${\frac{3}{2}}^{\pm}$ baryon states $B^*_{\pm}$  and the
${\frac{1}{2}}^{\pm}$ baryon states $B_{\pm}$ \cite{Chung82}, i.e.
\begin{eqnarray}
\langle{0}|J^{+}_{\mu}(0)| B_{\pm}^*(p)\rangle \langle
B_{\pm}^*(p)|\bar{J}^{+}_{\nu}(0)|0\rangle &=&
    - \gamma_{5}\langle 0|J^{-}_{\mu}(0)| B_{\pm}^*(p)\rangle \langle B_{\pm}^*(p)| \bar{J}^{-}_{\nu}(0)|0\rangle \gamma_{5} \,
, \nonumber \\
    \langle{0}|J^{+}_{\mu}(0)| B_{\pm}(p)\rangle \langle B_{\pm}(p)|\bar{J}^{+}_{\nu}(0)|0\rangle
&=&    - \gamma_{5}\langle 0|J^{-}_{\mu}(0)| B_{\pm}(p)\rangle
\langle B_{\pm}(p)| \bar{J}^{-}_{\nu}(0)|0\rangle \gamma_{5} \, ,
\end{eqnarray}
where
\begin{eqnarray}
\langle 0| J^{\pm}_\mu (0)|B_{\pm}^*(p)\rangle &=&\lambda_{\pm} U_\mu(p,s) \, , \nonumber \\
 \langle0|J^{\pm}_{\mu}(0)|B^{\mp}(p)\rangle&=&\lambda_{*}
 \left(\gamma_{\mu}-4\frac{p_{\mu}}{M_{*}}\right)U(p,s) \, ,
\end{eqnarray}
the $\lambda_{\pm}$ and $\lambda^{*}$ are  the  pole residues  and
$M_{*}$ are the masses, and  the spinor $U(p,s)$  satisfies the
usual Dirac equation $(\not\!\!p-M_{*})U(p)=0$.

The $\Pi^{\pm}_{\mu\nu}(p)$ have the following relation
\begin{eqnarray}
   \Pi^{-}_{\mu\nu}(p) &=& -\gamma_{5} \Pi^{+}_{\mu\nu}(p)\gamma_{5}   \, .
\end{eqnarray}

We  insert  a complete set  of intermediate baryon states with the
same quantum numbers as the current operators $J_\mu^{\pm}(x)$  into
the correlation functions $\Pi^{+}_{\mu\nu}(p)$  to obtain the
hadronic representation \cite{SVZ79,PRT85}. After isolating the pole
terms of the lowest states of the heavy and doubly heavy baryons, we
obtain the following result \cite{Oka96}:
\begin{eqnarray}
    \Pi^{+}_{\mu\nu}(p)     & = & -  \lambda_+^2 {\!\not\!{p} +
    M_{+} \over M^{2}_+ -p^{2} }g_{\mu\nu} - \lambda_{-}^2
    {\!\not\!{p} - M_{-} \over M_{-}^{2}-p^{2}  } g_{\mu\nu} +\cdots \,
, \nonumber \\
&=&-\Pi_+(p)g_{\mu\nu}+\cdots \, ,
    \end{eqnarray}
where the $M_{\pm}$ are the masses of the lowest states with parity
$\pm$ respectively, and the $\lambda_{\pm}$ are the  corresponding
pole residues (or couplings). In calculations, we have used the
following equations,
\begin{eqnarray}
\sum_s U_\mu(p,s) \overline{U}_\nu(p,s)
&=&(\!\not\!{p}+M_{B^*})\left( -g_{\mu\nu}+\frac{\gamma_\mu
\gamma_\nu}{3}+\frac{2p_\mu p_\nu}{3M_{B^*}^2}-\frac{p_\mu
\gamma_\nu-p_\nu \gamma_\mu}{3M_{B^*}} \right) \,  , \nonumber \\
\sum_sU(p,s)\overline {U}(p,s)&=&\!\not\!{p}+M_{*} \, .
\end{eqnarray}
In this article, we choose the tensor structure $g_{\mu\nu}$ for
analysis, the ${1\over 2}^\pm$ baryon states have no contaminations.

 If we take $\vec{p} = 0$, then
\begin{eqnarray}
  \rm{limit}_{\epsilon\rightarrow0}\frac{{\rm Im}  \Pi_+(p_{0}+i\epsilon)}{\pi} & = &
    \lambda_+^2 {\gamma_{0} + 1\over 2} \delta(p_{0} - M_+) +
    \lambda_{-}^{2} {\gamma_{0} - 1\over 2} \delta(p_{0} - M_{-})+\cdots \nonumber \\
  & = & \gamma_{0} A(p_{0}) + B(p_{0})+\cdots \, ,
\end{eqnarray}
where
\begin{eqnarray}
  A(p_{0}) & = & {1 \over 2} \left[ \lambda_+^{2}
  \delta(p_{0} - M_+)  + \lambda_-^{2} \delta(p_{0} -
  M_{-})\right] \, , \nonumber \\
   B(p_{0}) & = & {1 \over 2} \left[ \lambda_+^{2}
  \delta(p_{0} - M_+)  - \lambda_-^{2} \delta(p_{0} -
  M_{-})\right] \, ,
\end{eqnarray}
the  $A(p_{0}) + B(p_{0})$ and $A(p_{0}) - B(p_{0})$ contain the
contributions  from the positive-parity states and negative-parity
baryon states respectively.

We  calculate the light quark parts of the correlation functions
$\Pi^{+}_{\mu\nu}(p)$ in the coordinate space and use the momentum
space expression for the heavy quark propagators, i.e. we take
\begin{eqnarray}
S_{ij}(x)&=& \frac{i\delta_{ij}\!\not\!{x}}{ 2\pi^2x^4}
-\frac{\delta_{ij}m_s}{4\pi^2x^2}-\frac{\delta_{ij}}{12}\langle
\bar{s}s\rangle +\frac{i\delta_{ij}}{48}m_s
\langle\bar{s}s\rangle\!\not\!{x}     \nonumber\\
&& -\frac{i}{32\pi^2x^2}  G^{ij}_{\mu\nu}(x) \left[\!\not\!{x}
\sigma^{\mu\nu}+\sigma^{\mu\nu} \!\not\!{x}\right]  +\cdots \, ,\nonumber\\
S_Q^{ij}(x)&=&\frac{i}{(2\pi)^4}\int d^4k e^{-ik \cdot x} \left\{
\frac{\delta_{ij}}{\!\not\!{k}-m_Q}
-\frac{g_sG^{\alpha\beta}_{ij}}{4}\frac{\sigma_{\alpha\beta}(\!\not\!{k}+m_Q)+(\!\not\!{k}+m_Q)
\sigma_{\alpha\beta}}{(k^2-m_Q^2)^2}\right.\nonumber\\
&&\left.+\frac{\pi^2}{3} \langle \frac{\alpha_sGG}{\pi}\rangle
\delta_{ij}m_Q \frac{k^2+m_Q\!\not\!{k}}{(k^2-m_Q^2)^4}
+\cdots\right\} \, ,
\end{eqnarray}
where $\langle \frac{\alpha_sGG}{\pi}\rangle=\langle
\frac{\alpha_sG_{\alpha\beta}G^{\alpha\beta}}{\pi}\rangle$, then
resort to the Fourier integral to transform  the light quark parts
into the momentum space in $D$ dimensions,  take $\vec{p} = 0$,  and
use the dispersion relation to obtain the spectral densities
$\rho^A(p_0)$ and $\rho^B(p_0)$ (which correspond to the tensor
structures $\gamma_0$ and $1$ respectively) at the level of
quark-gluon degrees of freedom. Finally we introduce the weight
functions $\exp\left[-\frac{p_0^2}{T^2}\right]$,
$p_0^2\exp\left[-\frac{p_0^2}{T^2}\right]$,   and obtain the
following sum rules,
\begin{eqnarray}
  \lambda_{+}^2\exp\left[-\frac{M_+^2}{T^2}\right]&=&\int_{\Delta}^{\sqrt{s_0}}dp_0
\left[\rho^A(p_0)
+\rho^B(p_0)\right]\exp\left[-\frac{p_0^2}{T^2}\right] \, ,
\end{eqnarray}
\begin{eqnarray}
  \lambda_{+}^2M_+^2\exp\left[-\frac{M_+^2}{T^2}\right]&=&\int_{\Delta}^{\sqrt{s_0}}dp_0
\left[\rho^A(p_0)
+\rho^B(p_0)\right]p_0^2\exp\left[-\frac{p_0^2}{T^2}\right] \, ,
\end{eqnarray}
where  the $s_0$ are the threshold parameters, $T^2$ are the Borel
parameters, and $\Delta=2m_Q+m_s$,  $2m_Q$, $m_Q+2m_s$, $m_Q+m_s$
and $m_Q$ in the channels $\Omega^*_{QQ}$, $\Xi^*_{QQ}$,
$\Omega^*_Q$, $\Xi^*_Q$ and $\Sigma_Q^*$ respectively. The spectral
densities $\rho^A(p_0)$ and $\rho^B(p_0)$ at the level of
quark-gluon degrees of freedom are given explicitly in the Appendix.

\section{Numerical results and discussions}
The input parameters are taken to be the standard values $\langle
\bar{q}q \rangle=-(0.24\pm 0.01 \,\rm{GeV})^3$,  $\langle \bar{s}s
\rangle=(0.8\pm 0.2 )\langle \bar{q}q \rangle$, $\langle
\bar{q}g_s\sigma Gq \rangle=m_0^2\langle \bar{q}q \rangle$, $\langle
\bar{s}g_s\sigma Gs \rangle=m_0^2\langle \bar{s}s \rangle$,
$m_0^2=(0.8 \pm 0.2)\,\rm{GeV}^2$ \cite{Ioffe2005,LCSRreview},
$\langle \frac{\alpha_s GG}{\pi}\rangle=(0.012 \pm
0.004)\,\rm{GeV}^4 $ \cite{LCSRreview},
$m_s=(0.14\pm0.01)\,\rm{GeV}$, $m_c=(1.35\pm0.10)\,\rm{GeV}$ and
$m_b=(4.7\pm0.1)\,\rm{GeV}$ \cite{PDG} at the energy scale  $\mu=1\,
\rm{GeV}$.

The value of the gluon condensate $\langle \frac{\alpha_s
GG}{\pi}\rangle $ has been updated from time to time, and changes
greatly \cite{NarisonBook}.
 At the present case, the gluon condensate  makes tiny  contribution,
the updated value $\langle \frac{\alpha_s GG}{\pi}\rangle=(0.023 \pm
0.003)\,\rm{GeV}^4 $ \cite{NarisonBook} and the standard value
$\langle \frac{\alpha_s GG}{\pi}\rangle=(0.012 \pm
0.004)\,\rm{GeV}^4 $ \cite{LCSRreview} lead to a tiny  difference
and can be neglected safely.
 The values of the quark condensates determined from the
Gell-Mann-Oakes-Renner relation,  the spectral functions of the
$\tau$ decay, and the QCD sum rules for baryon masses are consistent
with each other within uncertainties \cite{Ioffe2005},  we usually
take the value from the Gell-Mann-Oakes-Renner relation in the QCD
sum rules \cite{LCSRreview}. For the mixed condensates, we  take the
value  from the QCD sum rules for the baryonic resonances, which is
still accepted in the literatures \cite{Ioffe2005,LCSRreview}. Those
values are not accurate, and there are much room for improvement;
the update of the vacuum condensates should be combined with a more
delicate procedure in dealing with the  perturbative and
non-perturbative contributions, and beyond the present work.

The $Q$-quark masses appearing in the perturbative terms  are
usually taken to be the pole masses in the QCD sum rules, while the
choice of the $m_Q$ in the leading-order coefficients of the
higher-dimensional terms is arbitrary \cite{NarisonBook,Kho9801}.
The $\overline{MS}$ mass $m_c(m_c^2)$ relates with the pole mass
$\hat{m}$ through the relation $ m_c(m_c^2)
=\hat{m}\left[1+\frac{C_F \alpha_s(m_c^2)}{\pi}+\cdots\right]^{-1}
$. In this article, we take the approximation $m_c\approx\hat{m}$
without the $\alpha_s$ corrections for consistency. The value listed
in the Particle Data Group is $m_c(m_c^2)=1.27^{+0.07}_{-0.11} \,
\rm{GeV}$ \cite{PDG}, it is reasonable to take
$m_c=m_c(1\,\rm{GeV}^2)=(1.35\pm0.10)\,\rm{GeV}$. The value of the
 $m_b$ can be understood   analogously.

In calculation, we  also neglect  the contributions from the
perturbative $\mathcal {O}(\alpha_s^n)$  corrections.  Those
perturbative corrections can be taken into account in the leading
logarithmic  approximations through the anomalous dimension factors.
After the Borel transform, the effects of those
 corrections are  to multiply each term on the operator product
 expansion side by the factor, $ \left[ \frac{\alpha_s(T^2)}{\alpha_s(\mu^2)}\right]^{2\Gamma_{J}-\Gamma_{\mathcal
 {O}_n}}  $,
 where the $\Gamma_{J}$ is the anomalous dimension of the
 interpolating current $J(x)$, and the $\Gamma_{\mathcal {O}_n}$ is the anomalous dimension of
 the local operator $\mathcal {O}_n(0)$, which
governs the evolution of the vacuum condensate
$\langle{O}_n(0)\rangle_\mu$ with the energy scale through the
re-normalization group equation.

 If the perturbative
$\mathcal {O}(\alpha_s)$ corrections and the anomalous dimension
factors are taken into account consistently, the spectral densities
in the QCD side should be replaced with
\begin{eqnarray}
\mathcal {O}_0(0) &\rightarrow &\left[
\frac{\alpha_s(T^2)}{\alpha_s(\mu^2)}\right]^{2\Gamma_{J}}
\left[1+A(p_0^2,m_Q^2)\frac{\alpha_s(T^2)}{\pi} \right]\mathcal
{O}_0(0) \, ,\nonumber \\
\langle\mathcal {O}_n(0)\rangle_{\mu} &\rightarrow &\left[
\frac{\alpha_s(T^2)}{\alpha_s(\mu^2)}\right]^{2\Gamma_{J}-\Gamma_{\mathcal
 {O}_n} }\left[1+B(p_0^2,m_Q^2)\frac{\alpha_s(T^2)}{\pi} \right]\langle\mathcal{O}_n(0)\rangle_{\mu} \, ,\nonumber
\end{eqnarray}
where the $A(p_0^2,m_Q^2)$ and $B(p_0^2,m_Q^2)$ are some notations
for the coefficients of the perturbative corrections, the average
virtuality of the quarks in the correlation functions is
 characterized by the Borel parameter $T^2$. We cannot estimate the
corrections and the uncertainties originate from the corrections
with confidence without explicit calculations. In Ref.\cite{Sulian},
Ovchinnikov et al calculate the perturbative $\mathcal
{O}(\alpha_s)$ corrections to  the correlation functions of the
light-flavor  baryon, and observe that the corrections change
 the numerical values of the mass and the pole residue
of the proton considerably and improve the agreement between the
theoretical estimation and the experimental data. In the present
case, including the $\alpha_s$ corrections maybe improve the
predictions.

In this article, we carry out the operator product expansion at the
special energy scale $\mu^2=1\,\rm{GeV}^2$, and  set the factor
$\left[\frac{\alpha_s(T^2)}{\alpha_s(\mu^2)}\right]^{2\Gamma_{J}-\Gamma_{\mathcal
{O}_n}}\approx1$ for consistency, as the $\alpha_s$ corrections have
not been calculated yet. Such an approximation maybe result in some
scale dependence and weaken the prediction ability. In this article,
we study the $\frac{3}{2}^+$ heavy and doubly heavy baryon states in
a systematic  way, the predictions are still robust   as we take the
analogous criteria in those sum rules.

The separation of the perturbative and non-perturbative
contributions to the vacuum correlation functions has some
arbitrariness, and we can introduce some renormalization point $\mu$
($\mu^2\sim 1\,\rm{GeV}^2$) as the boundary. The non-perturbative
contributions are parameterized by the vacuum condensates,
furthermore, the infrared logarithms of the form $\log^k\left(
\frac{m_q^2}{\mu^2}\right)$ are also absorbed into the vacuum
condensates in the perturbative calculations. Perturbative
calculations are reliable at the special energy scale
$\mu^2=1\,\rm{GeV}^2$, which characterizes the chiral symmetry
breaking.

\begin{figure}
 \centering
 \includegraphics[totalheight=4cm,width=5cm]{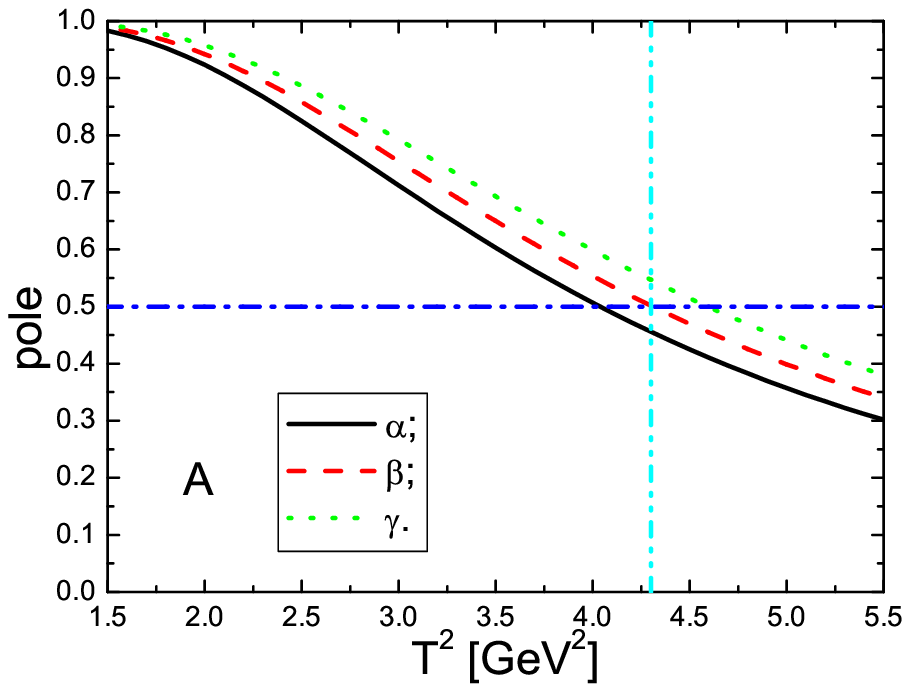}
 \includegraphics[totalheight=4cm,width=5cm]{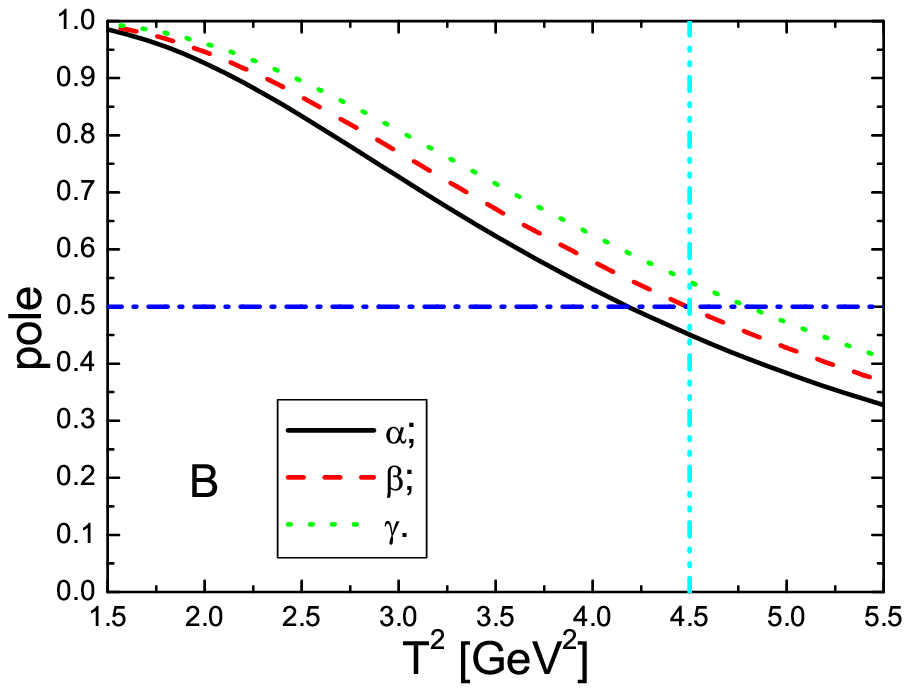}
\includegraphics[totalheight=4cm,width=5cm]{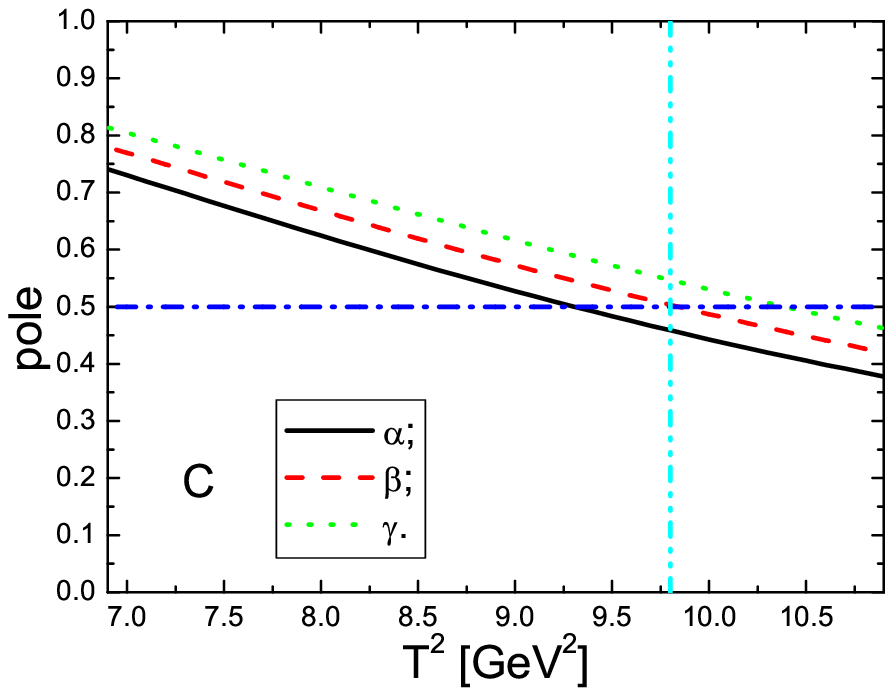}
\includegraphics[totalheight=4cm,width=5cm]{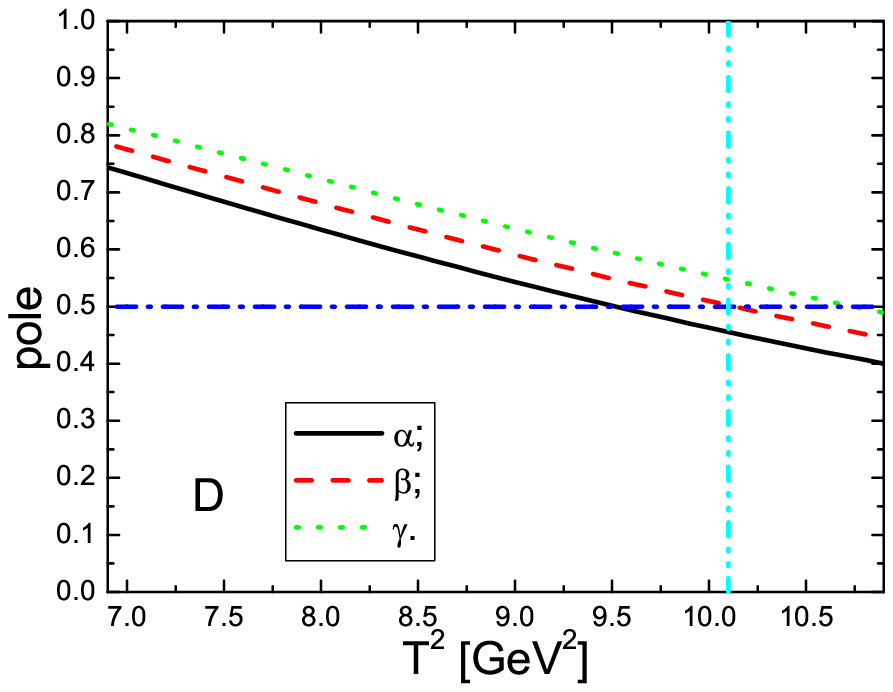}
\includegraphics[totalheight=4cm,width=5cm]{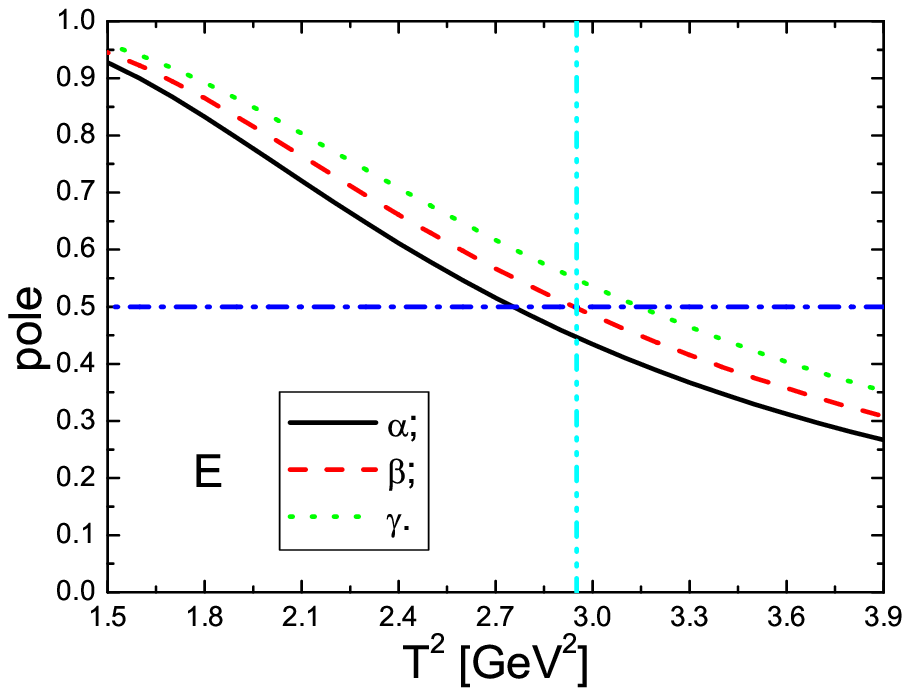}
 \includegraphics[totalheight=4cm,width=5cm]{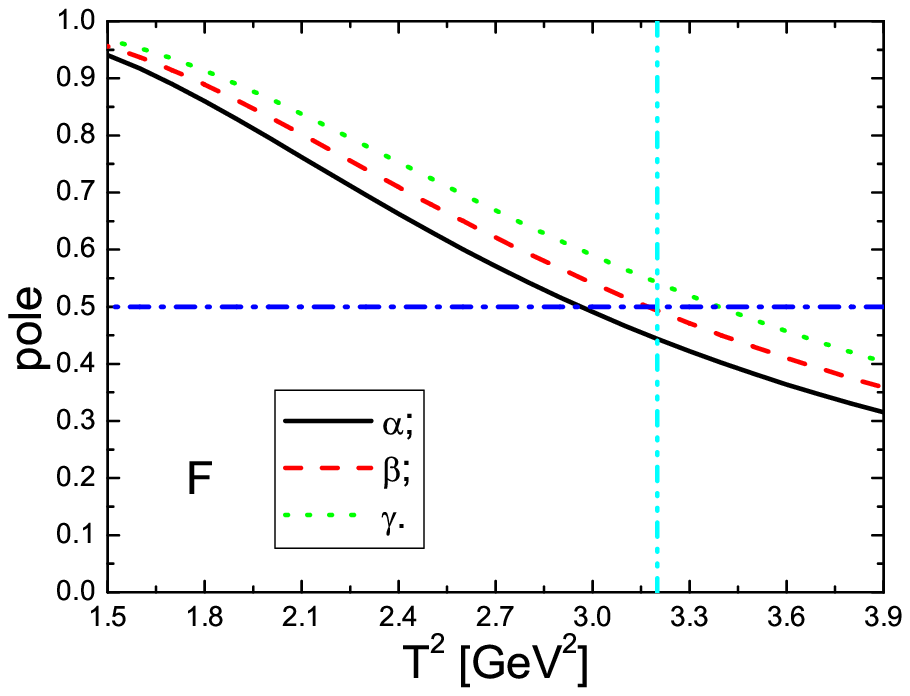}
\includegraphics[totalheight=4cm,width=5cm]{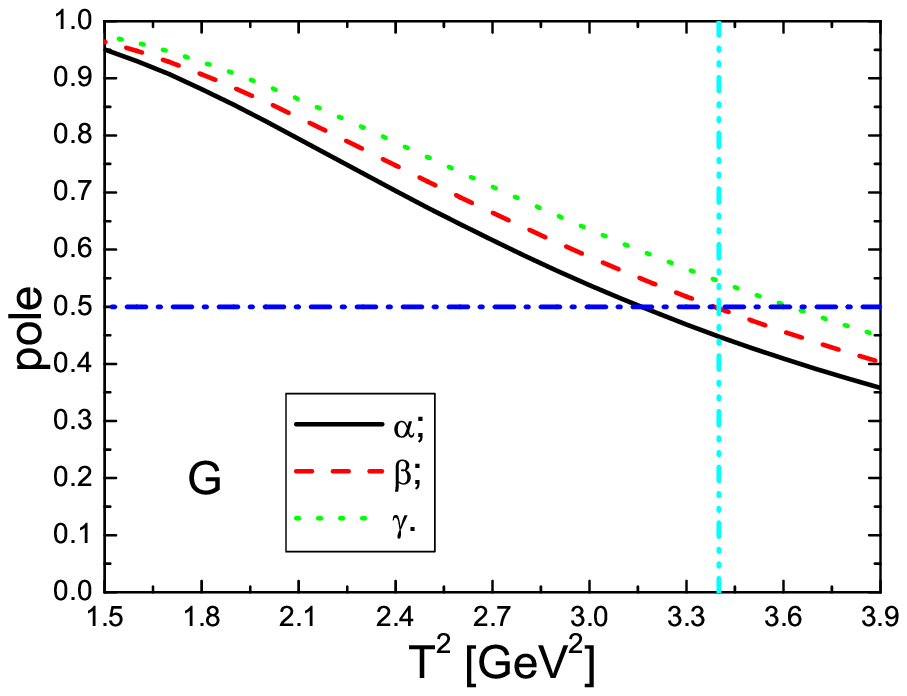}
\includegraphics[totalheight=4cm,width=5cm]{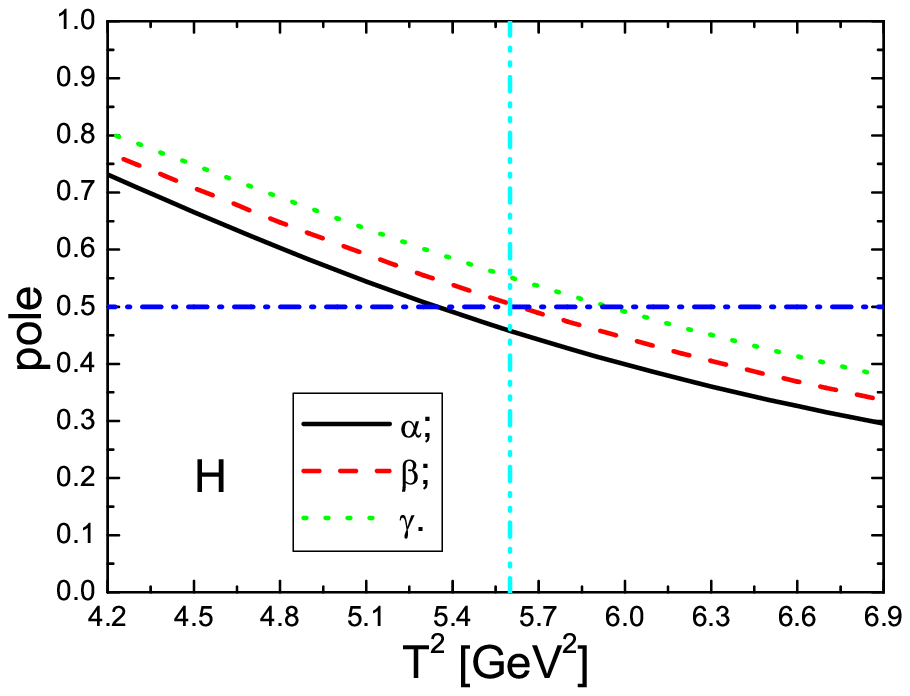}
\includegraphics[totalheight=4cm,width=5cm]{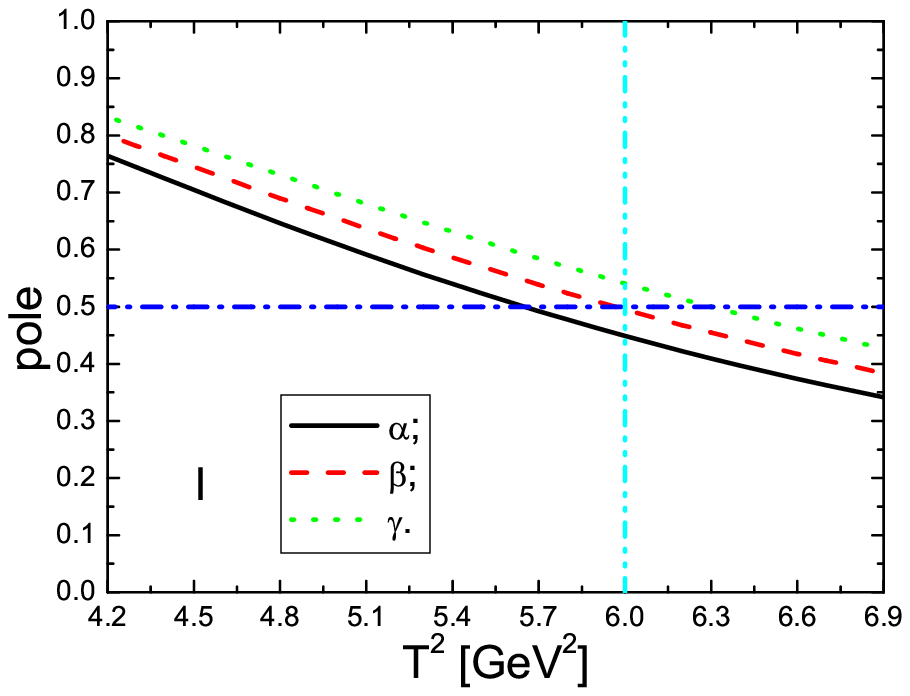}
 \includegraphics[totalheight=4cm,width=5cm]{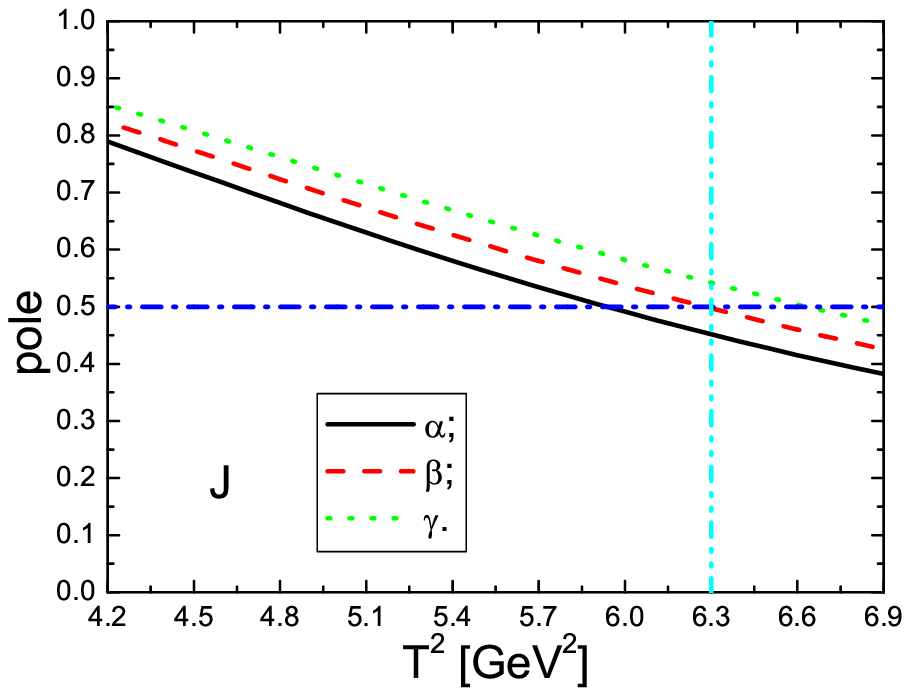}
        \caption{ The contributions of the pole terms with variations of the Borel parameters $T^2$, the
$A$, $B$, $C$, $D$, $E$, $F$, $G$, $H$, $I$ and $J$  correspond
      to the channels $\Xi^*_{cc}$, $\Omega^*_{cc}$, $\Xi^*_{bb}$,  $\Omega^*_{bb}$, $\Sigma_c^*$,
$\Xi_c^*$, $\Omega_c^*$, $\Sigma_b^*$, $\Xi_b^*$ and $\Omega_b^*$
respectively, the $\beta$ corresponds to the central values of the
threshold parameters, the energy gap among $\alpha$, $\beta$ and
$\gamma$ is $0.1\,\rm{GeV}$. }
\end{figure}

\begin{figure}
 \centering
 \includegraphics[totalheight=4cm,width=5cm]{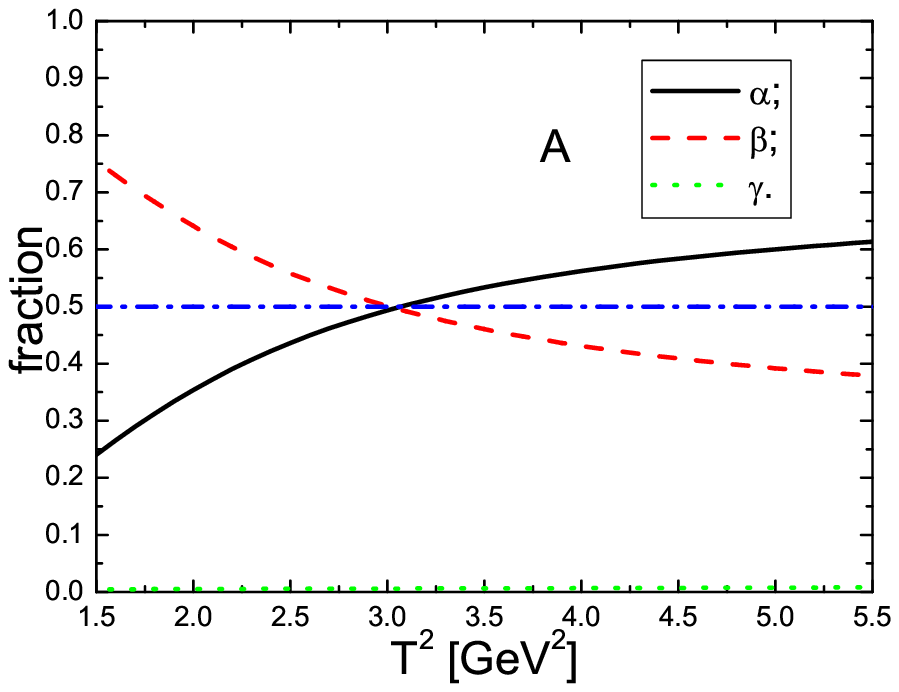}
 \includegraphics[totalheight=4cm,width=5cm]{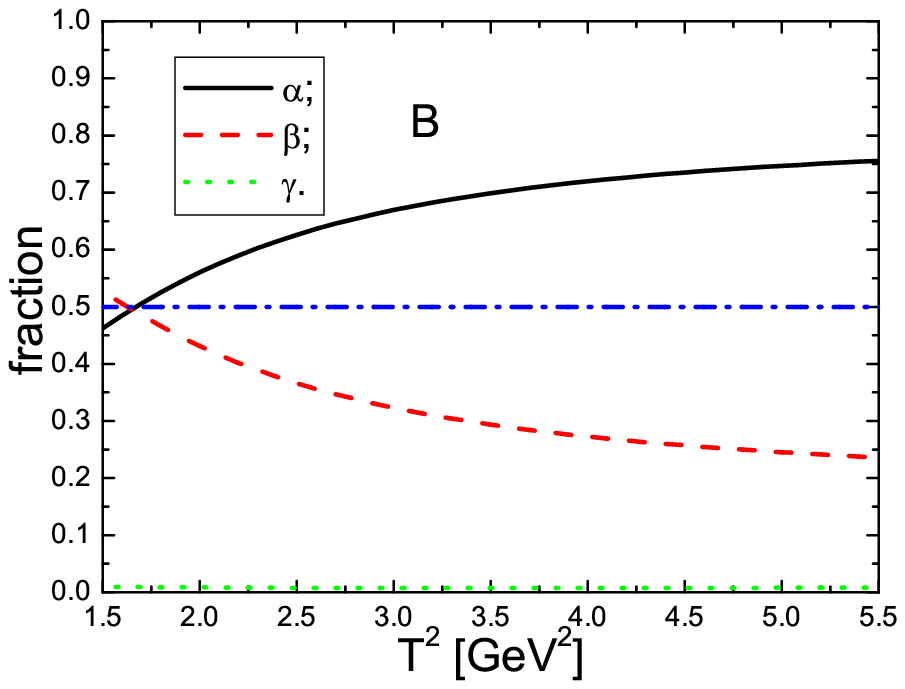}
\includegraphics[totalheight=4cm,width=5cm]{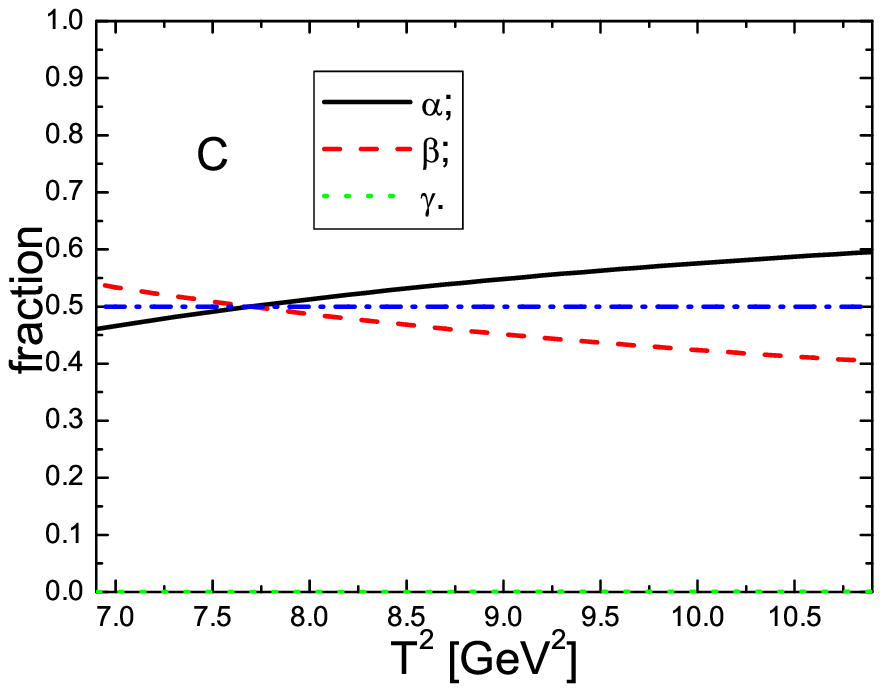}
\includegraphics[totalheight=4cm,width=5cm]{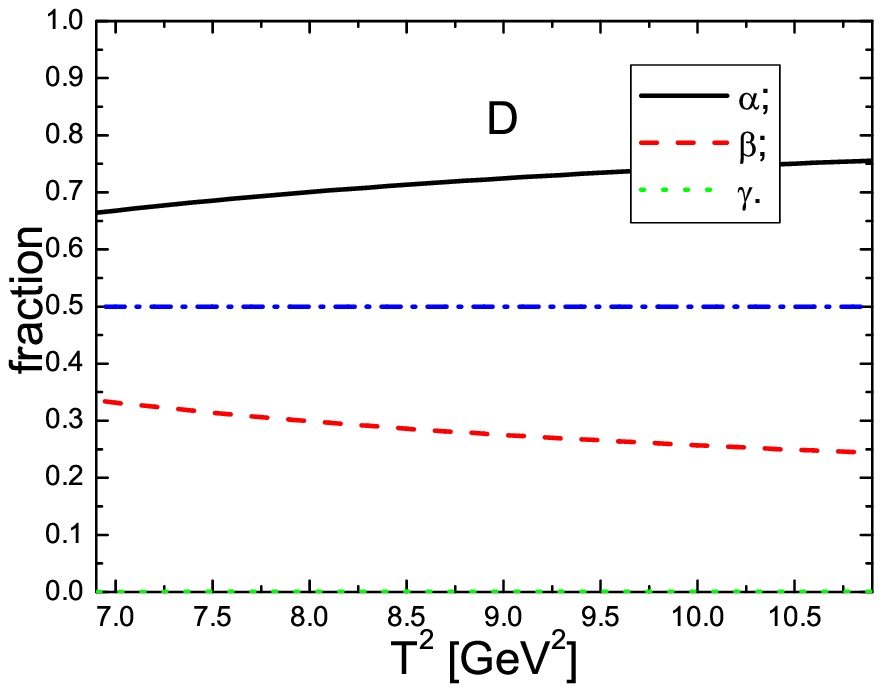}
        \caption{ The contributions from different terms in the operator product expansion, the $A$, $B$, $C$ and $D$  correspond
      to the channels $\Xi^*_{cc}$, $\Omega^*_{cc}$, $\Xi^*_{bb}$ and $\Omega^*_{bb}$ respectively, the $\alpha$, $\beta$ and $\gamma$ denote
the perturbative term, the quark condensate term and the gluon
condensate term respectively, the threshold parameters are taken to
be the central values. }
\end{figure}

In the conventional QCD sum rules \cite{SVZ79,PRT85}, there are two
criteria (pole dominance and convergence of the operator product
expansion) for choosing  the Borel parameter $T^2$ and threshold
parameter $s_0$.  We impose the two criteria on the heavy and doubly
heavy baryon states to choose the Borel parameter $T^2$ and
threshold parameter $s_0$.

In Fig.1, we plot the  contributions from the pole terms with
variations  of the Borel parameters $T^2$ and the threshold
parameters $s_0$. The pole contributions are larger than (or about)
$50\%$ at the values which are denoted by the vertical lines for
central values ($\beta$) of the threshold parameters $s_0$. We can
set the upper bound of the  Borel parameters $T^2_{max}$ as the
values indicated by the vertical lines.

  In Fig.2, we plot the
contributions from the different terms in the operator product
expansion in the doubly heavy baryon channels $\Xi^*_{cc}$,
$\Omega^*_{cc}$, $\Xi^*_{bb}$ and $\Omega^*_{bb}$. In the heavy
baryon channels $\Sigma_c^*$, $\Xi_c^*$, $\Omega_c^*$, $\Sigma_b^*$,
$\Xi_b^*$ and $\Omega_b^*$, the convergent behaviors are very good,
it is no use to plot them,  we only show the contributions from the
perturbative terms explicitly in Table 1. From the Fig.2, we can see
that the convergent behaviors in the channels $\Omega^*_{QQ}$ are
better than the corresponding ones in the channels $\Xi^*_{QQ}$,
this is mainly due to the fact that the values of the quark
condensates, $|\langle \bar{q}{q}\rangle|>|\langle
\bar{s}{s}\rangle|$. The lower bound of the Borel parameters
$T^2_{min}$ can be determined by the channels $\Xi^*_{QQ}$ in the
regions  where the contributions from the perturbative terms are
larger than the corresponding ones from the quark condensates. From
Figs.1-2, we can see that the Borel windows are different for the
doubly charm and doubly bottom baryon states.

 In this article, we
take the uniform Borel windows, $T^2_{max}-T^2_{min}=1.5 \,
\rm{GeV}^2$ and $2.0 \, \rm{GeV}^2$ in the doubly charm  and doubly
bottom channels respectively. For the heavy baryon states
$\Sigma_c^*$, $\Xi_c^*$, $\Omega_c^*$, $\Sigma_b^*$, $\Xi_b^*$ and
$\Omega_b^*$, we take the uniform Borel windows,
$T^2_{max}-T^2_{min}=1.0 \, \rm{GeV}^2$. The values of the threshold
parameters $s_0$ and
   the Borel parameters $T^2$ are shown in Table 1, from the table,
    we can see that the two criteria of the QCD sum rules
are fully  satisfied  \cite{SVZ79,PRT85}. In this article, we take
uniform uncertainties for the threshold parameters,
$\delta_{\sqrt{s_0}}=\pm 0.1\, \rm{GeV}$. In calculation, we observe
that the   predicted masses are not sensitive to the threshold
parameters, although they increase with the threshold parameters.

Taking into account all uncertainties  of the revelent  parameters,
we can obtain the values of the masses and pole residues of
 the ${3\over 2}^+$ heavy and doubly heavy
baryon states $\Xi^*_{cc}$, $\Omega^*_{cc}$,  $\Xi^*_{bb}$,
$\Omega^*_{bb}$,
 $\Sigma_c^*$, $\Xi_c^*$, $\Omega_c^*$, $\Sigma_b^*$,
$\Xi_b^*$ and $\Omega_b^*$, which are shown in Figs.3-4 and Tables
2-4. In Table 2 and Table 4, we also present the predictions of
other theoretical approaches and the values of the experimental
data, respectively.

From Table 4, we can see that the present predictions for the well
established heavy baryon states are in good agreement with the
experimental data,   the predictions for the unestablished bottom
baryon states $\Xi_b^*$ and $\Omega_b^*$ are   robust as we take the
analogous criteria in those sum rules. For  the ${3\over 2}^+$
doubly heavy baryon states, there are no available experimental
data, our values  are comparable with other theoretical predictions,
see Table 2.

\begin{table}
\begin{center}
\begin{tabular}{|c|c|c|c|c|c|c|c|}
\hline\hline & $T^2 (\rm{GeV}^2)$& $\sqrt{s_0} (\rm{GeV})$&pole&perturbative& $\langle \bar{q}q\rangle$ & $\langle \frac{\alpha_sGG}{\pi}\rangle$\\
\hline
 $\Xi^*_{cc}$  &$2.8-4.3$ &$4.3$&  $(46-83)\%$&$(47-58)\%$ & $(42-52)\%$ &$<1\%$\\ \hline
   $\Omega^*_{cc}$  &$3.0-4.5$ &$4.4$& $(45-81)\%$&$(67-74)\%$& $(26-32)\%$ &$<1\%$\\ \hline
    $\Xi^*_{bb}$  &$7.8-9.8$ &$10.9$& $(46-73)\%$ &$(50-57)\%$& $(43-50)\%$&$\ll1\%$\\ \hline
    $\Omega^*_{bb}$  &$8.1-10.1$ &$11.0$ & $(46-71)\%$&$(70-74)\%$&$(26-30)\%$&$\ll1\%$\\ \hline\hline
$\Omega^*_b$  &$5.3-6.3$ &$6.9$&  $(45-68)\%$&$(84-89)\%$ &&\\
\hline
       $\Xi^*_b$  &$5.0-6.0$ &$6.8$& $(45-70)\%$&$(78-85)\%$&&\\ \hline
           $\Sigma^*_b$  &$4.6-5.6$ & $6.7$& $(46-73)\%$&$(71-83)\%$&&\\ \hline
           $\Omega^*_c$  &$2.4-3.4$ &$3.5$& $(45-79)\%$ &$(79-87)\%$&&\\ \hline
              $\Xi^*_c$  &$2.2-3.2$ &$3.4$ & $(44-81)\%$&$(72-85)\%$&&\\ \hline
              $\Sigma^*_c$  &$2.0-3.0$ &$3.3$& $(43-83)\%$&$(64-84)\%$&&\\ \hline
\hline
\end{tabular}
\end{center}
\caption{ The Borel parameters $T^2$ and threshold parameters $s_0$
for the heavy and doubly heavy baryon states, the "pole" stands for
the contribution from the pole term, and the "perturbative" stands
for the contribution from the perturbative term in the operator
product expansion, etc. In calculating the contributions from the
pole terms, we take into account the uniform uncertainties of  the
threshold parameters, $\delta_{\sqrt{s_0}}=\pm 0.1\, \rm{GeV}$.  }
\end{table}

\begin{table}
\begin{center}
\begin{tabular}{|c|c|c|c|c|}
\hline\hline Reference
&$\Xi^*_{cc}$&$\Omega^*_{cc}$&$\Xi^*_{bb}$&$\Omega^*_{bb}$\\ \hline
  \cite{HH-Valcarce}&  $3.656$  &$3.769$ &$10.218$ &$10.321$\\ \hline
\cite{HH-Roberts}& $3.753$ &$3.876$ &$10.367$ & $10.486$ \\ \hline
\cite{HH-Ebert}& $3.727$&$3.872$&$10.237$ & $10.389$\\ \hline
\cite{HH-Albertus}& $3.706$ &$3.783$ &$10.236$ & $10.297$\\ \hline
\cite{HH-Kiselev}& $3.61$ &$3.69$ &$10.13$ & $10.20$\\ \hline
 \cite{HH-Giannuzzi}&   $3.719$  &$3.770$ &$10.216$ & $10.289$\\ \hline
\cite{HH-He}& $3.630$ &$3.721$ &$10.337$ & $10.429$\\ \hline
 \cite{HH-Zhang}& $3.90$ &$3.81$ &$10.35$ & $10.28$\\ \hline
   This work&   $3.61\pm0.18$  &$3.76\pm0.17$ &$10.22\pm0.15$ & $10.38\pm0.14$\\ \hline\hline
\end{tabular}
\end{center}
\caption{ The masses $M(\rm{GeV})$   of the ${3\over 2}^+$ doubly
heavy baryon states.}
\end{table}

\begin{table}
\begin{center}
\begin{tabular}{|c|c|c|c|c|}
\hline\hline  & $\Xi^*_{cc}$& $\Omega^*_{cc}$&
$\Xi^*_{bb}$&$\Omega^*_{bb}$\\\hline
   $\lambda\,[\rm{GeV}^3]$&   $0.070\pm0.017$  &$0.085\pm0.019$ &$0.161\pm0.041$ & $0.199\pm0.048$\\ \hline\hline
\end{tabular}
\end{center}
\caption{ The  pole residues $\lambda$ of the ${3\over 2}^+$ doubly
 heavy baryon states.}
\end{table}

\begin{table}
\begin{center}
\begin{tabular}{|c|c|c|c|c|c|c|}
\hline\hline & $T^2 (\rm{GeV}^2)$& $\sqrt{s_0} (\rm{GeV})$&
$M(\rm{GeV})$&$\lambda (\rm{GeV}^3)$&$M(\rm{GeV})[\rm{exp}]$\\\hline
 $\Omega_b^*$  &$5.3-6.3$ &$6.9\pm0.1$&  $6.17\pm0.15$&$0.083\pm0.018$& ?\\ \hline
       $\Xi^*_b$  &$5.0-6.0$ &$6.8\pm0.1$& $6.02\pm0.17$&$0.049\pm0.012$&?\\ \hline
           $\Sigma^*_b$  &$4.6-5.6$ & $6.7\pm0.1$& $5.85\pm0.20$&$0.038\pm0.011$&5.833 \,\cite{PDG}\\ \hline
           $\Omega^*_c$  &$2.4-3.4$ &$3.5\pm0.1$& $2.79\pm0.19$&$0.056\pm0.012$ &2.766 \,\cite{PDG}\\ \hline
              $\Xi^*_c$  &$2.2-3.2$ &$3.4\pm0.1$ & $2.65\pm0.20$&$0.033\pm0.008$&2.646 \,\cite{PDG}\\ \hline
              $\Sigma^*_c$  &$2.0-3.0$ &$3.3\pm0.1$& $2.48\pm0.25$&$0.027\pm0.008$&2.518 \,\cite{PDG}\\ \hline
    \hline
\end{tabular}
\end{center}
\caption{ The masses $M(\rm{GeV})$ and pole residues
$\lambda(\rm{GeV}^3)$ of the ${3\over 2}^+$ heavy baryon states.}
\end{table}

\begin{figure}
 \centering
 \includegraphics[totalheight=4cm,width=5cm]{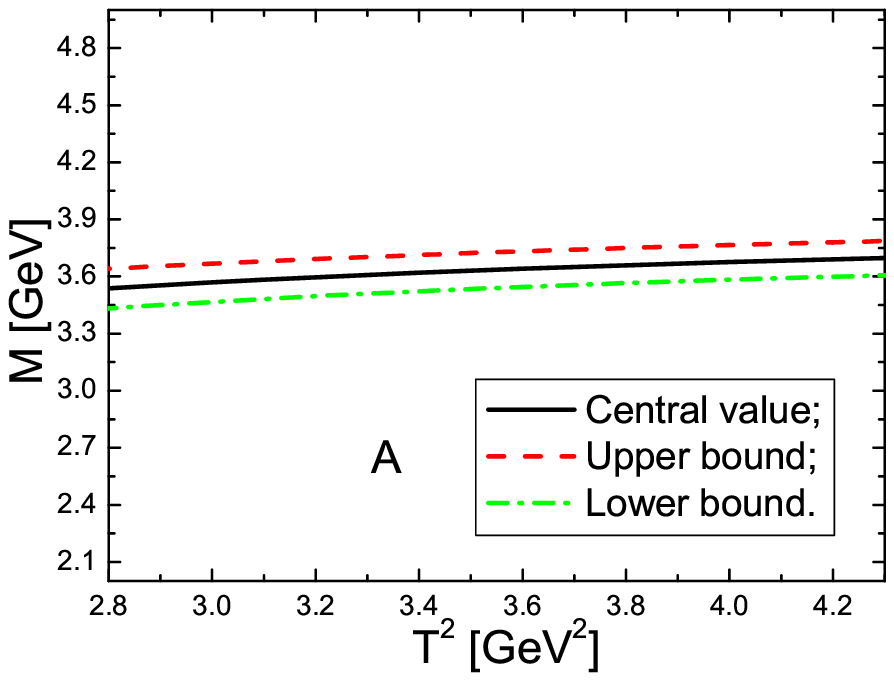}
 \includegraphics[totalheight=4cm,width=5cm]{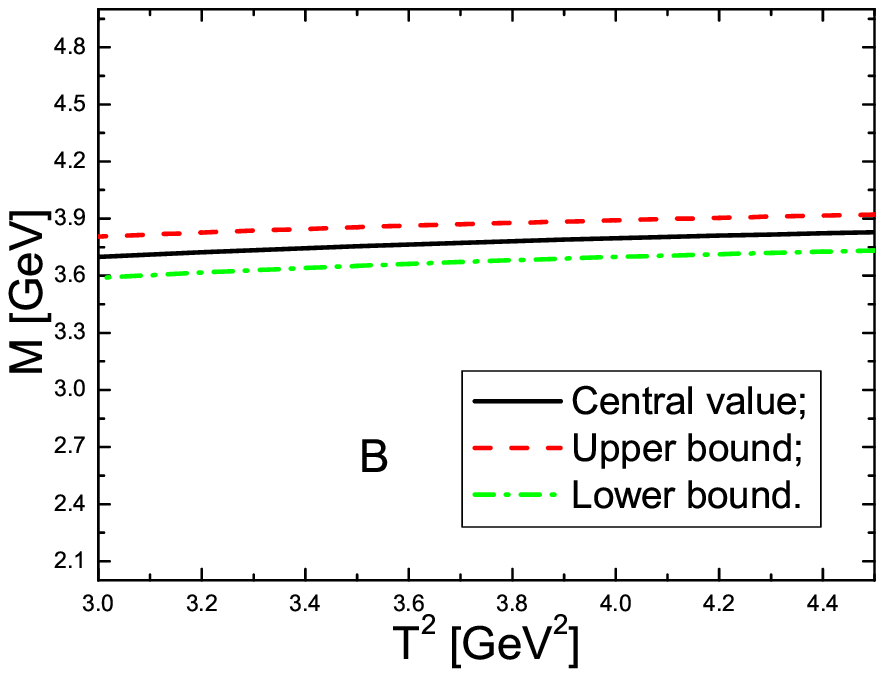}
\includegraphics[totalheight=4cm,width=5cm]{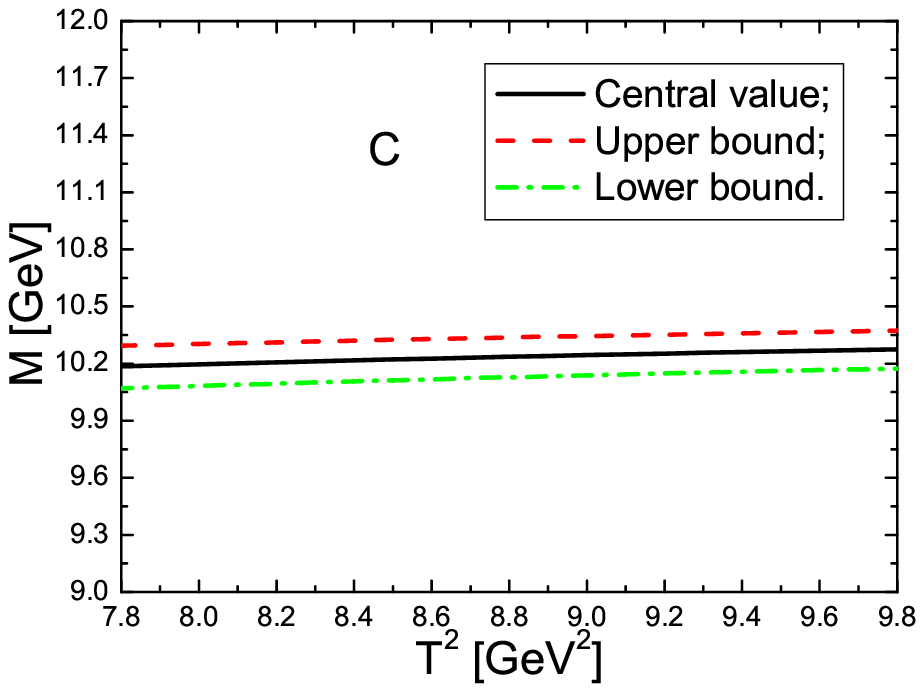}
\includegraphics[totalheight=4cm,width=5cm]{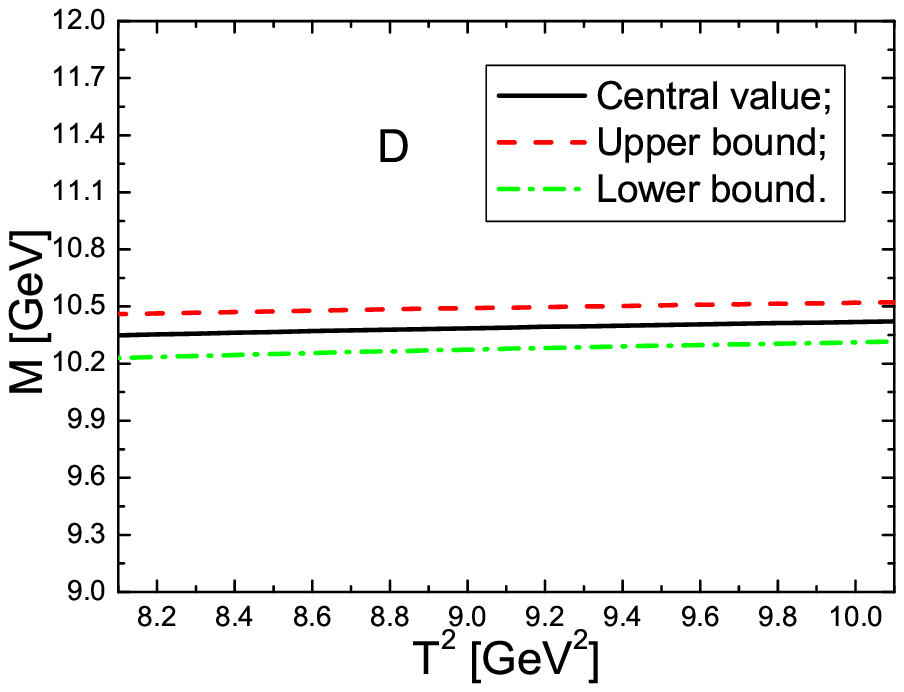}
\includegraphics[totalheight=4cm,width=5cm]{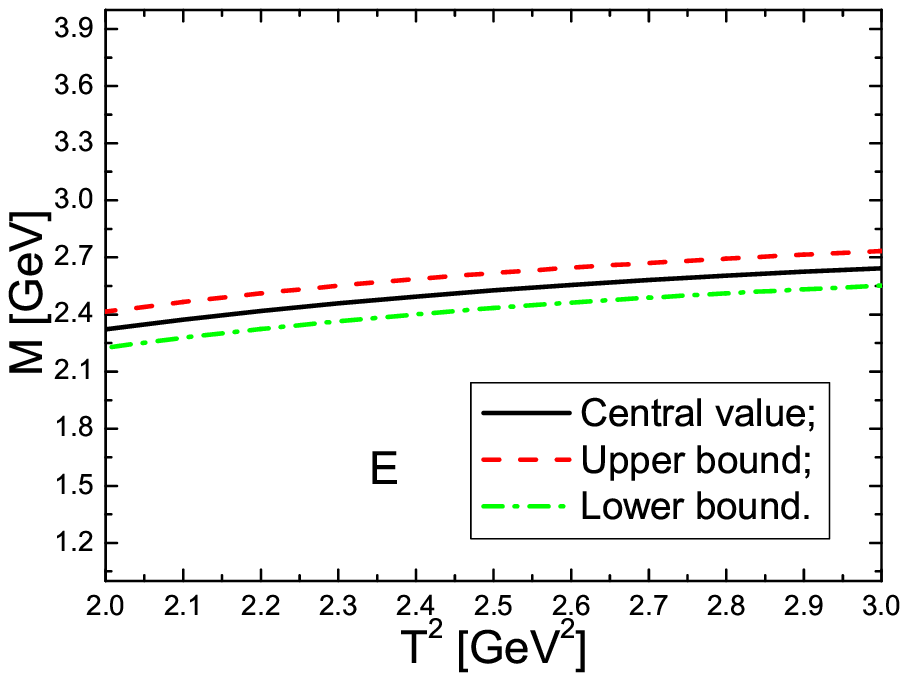}
 \includegraphics[totalheight=4cm,width=5cm]{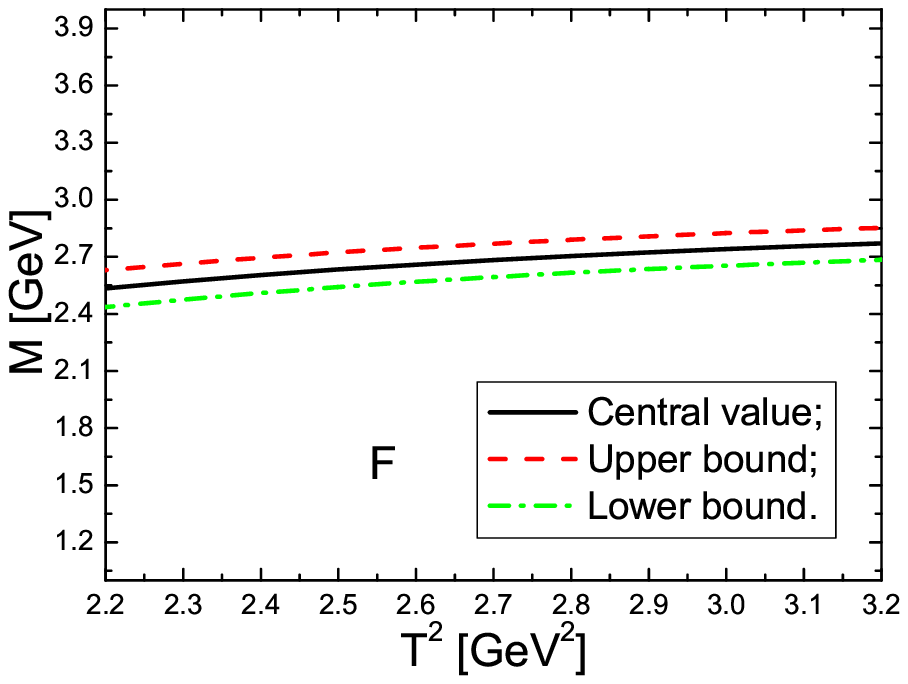}
\includegraphics[totalheight=4cm,width=5cm]{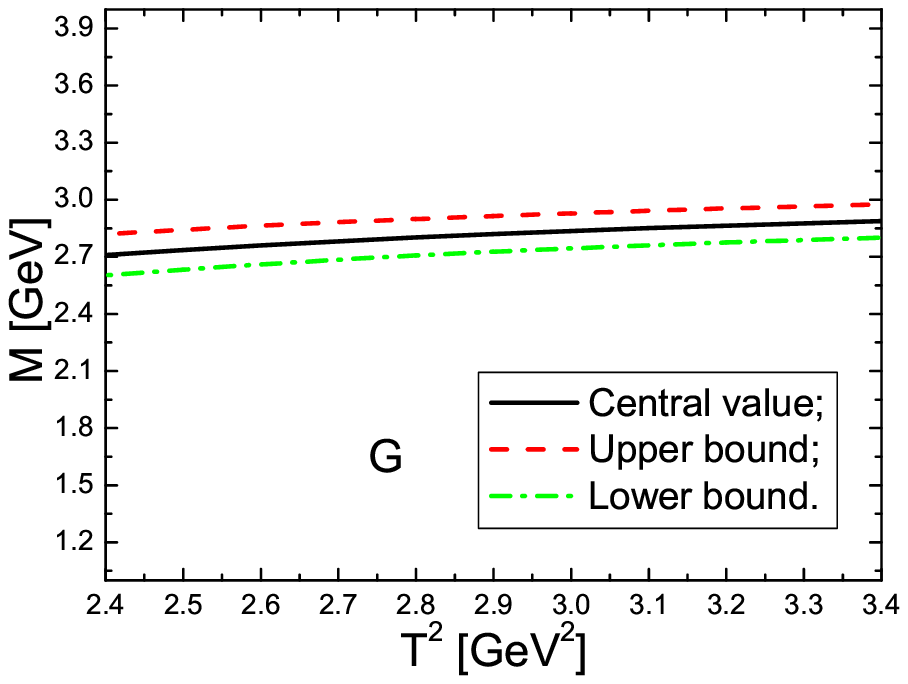}
\includegraphics[totalheight=4cm,width=5cm]{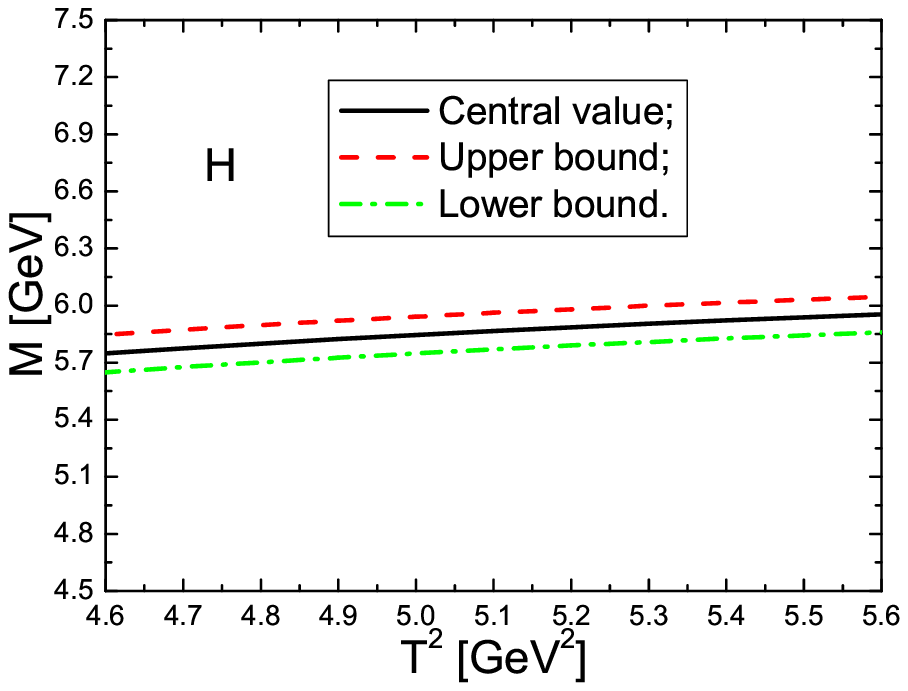}
\includegraphics[totalheight=4cm,width=5cm]{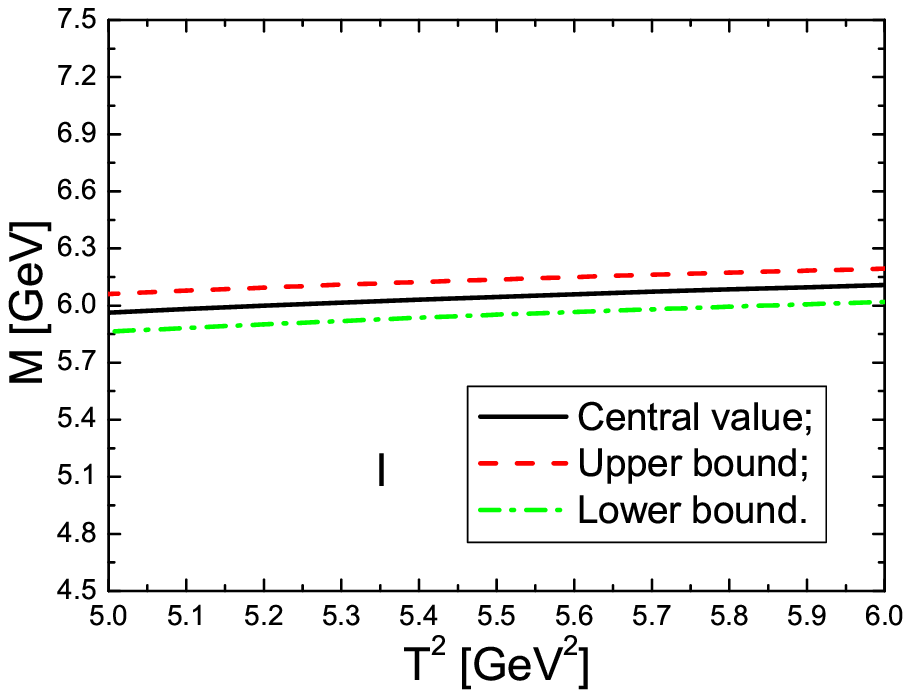}
 \includegraphics[totalheight=4cm,width=5cm]{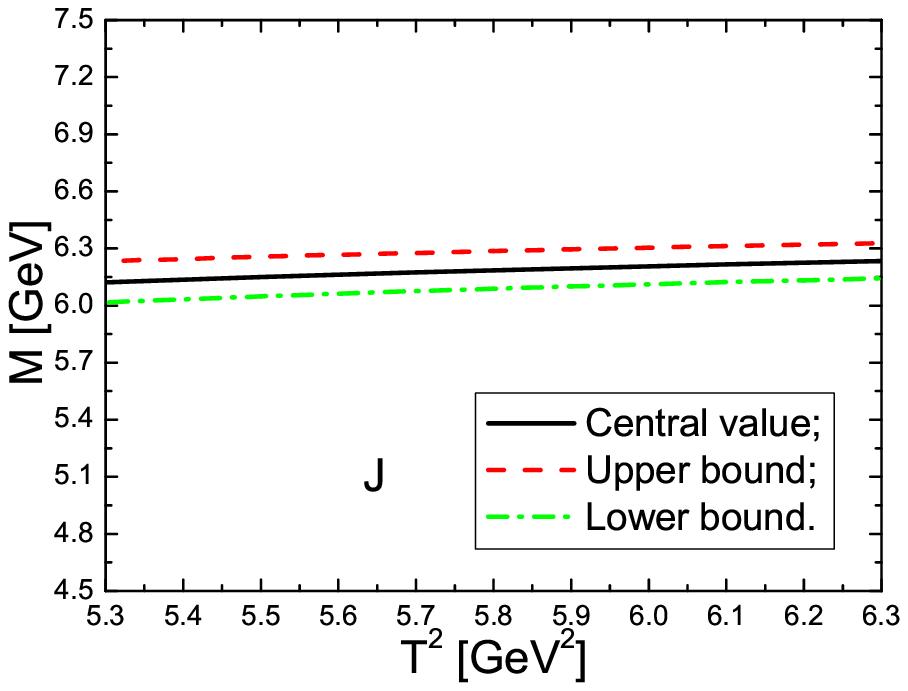}
        \caption{ The  masses of the heavy and doubly heavy baryon states, the
$A$, $B$, $C$, $D$, $E$, $F$, $G$, $H$, $I$ and $J$  correspond
      to the channels $\Xi^*_{cc}$, $\Omega^*_{cc}$, $\Xi^*_{bb}$,  $\Omega^*_{bb}$, $\Sigma_c^*$,
$\Xi_c^*$, $\Omega_c^*$, $\Sigma_b^*$, $\Xi_b^*$ and $\Omega_b^*$
respectively. }
\end{figure}

\begin{figure}
 \centering
 \includegraphics[totalheight=4cm,width=5cm]{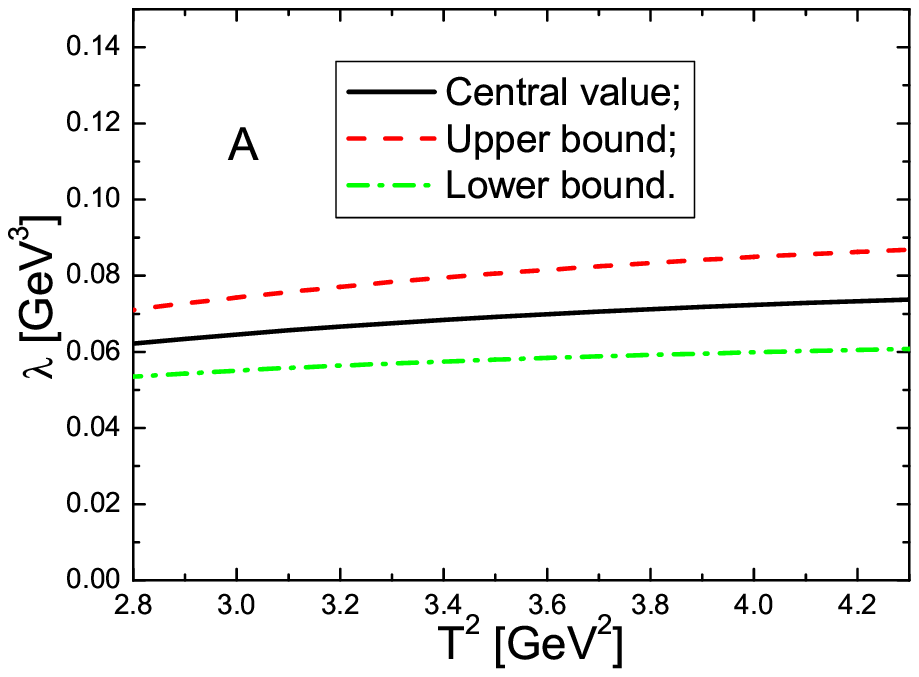}
 \includegraphics[totalheight=4cm,width=5cm]{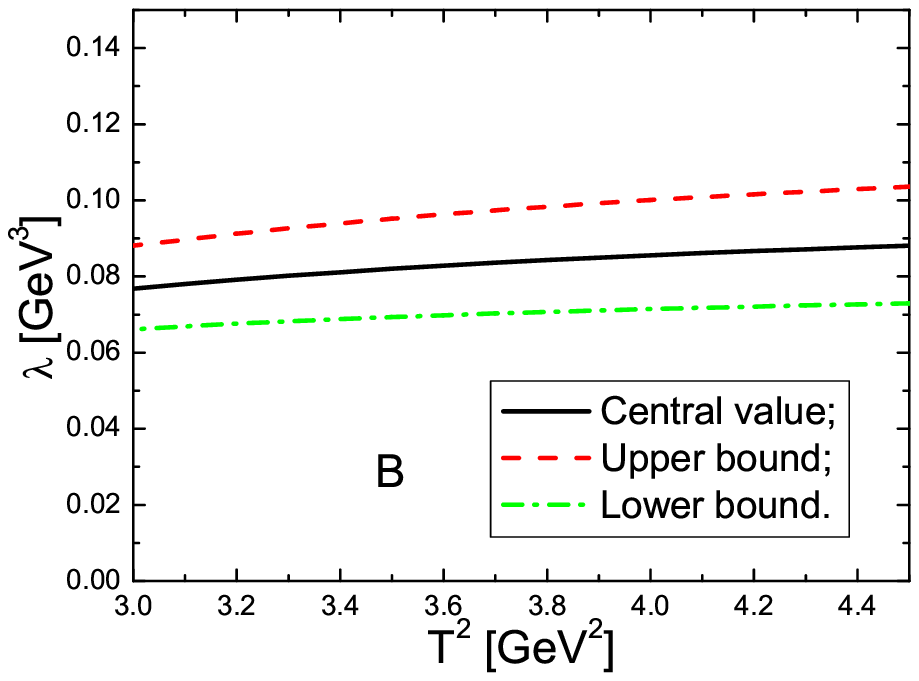}
\includegraphics[totalheight=4cm,width=5cm]{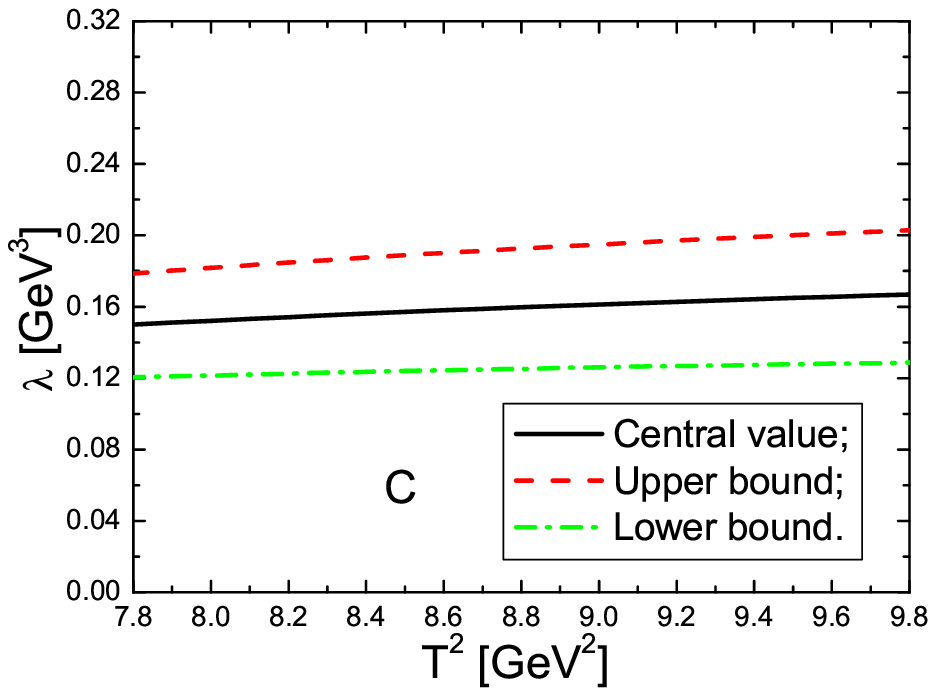}
\includegraphics[totalheight=4cm,width=5cm]{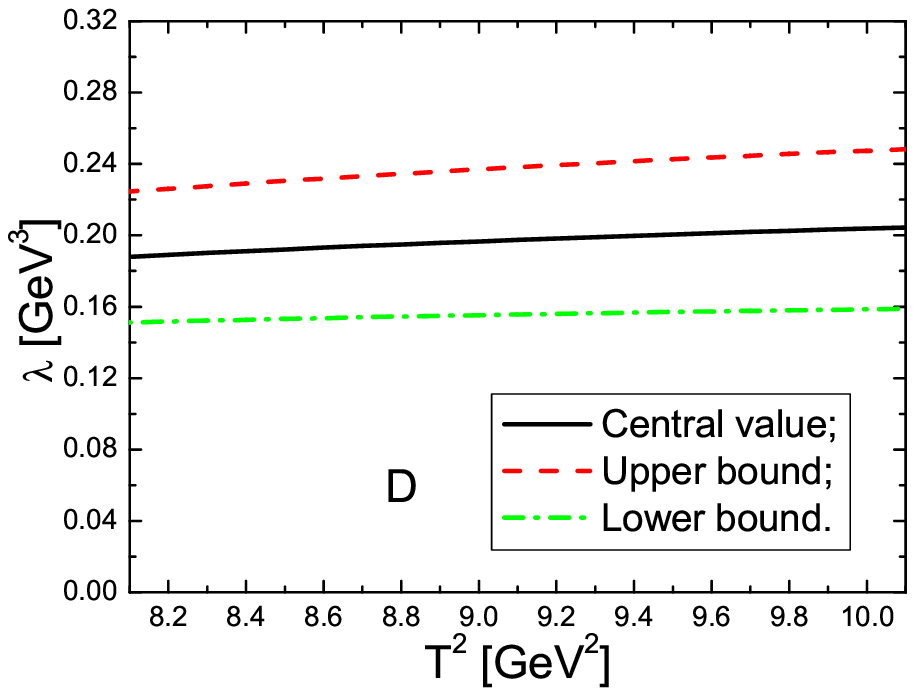}
\includegraphics[totalheight=4cm,width=5cm]{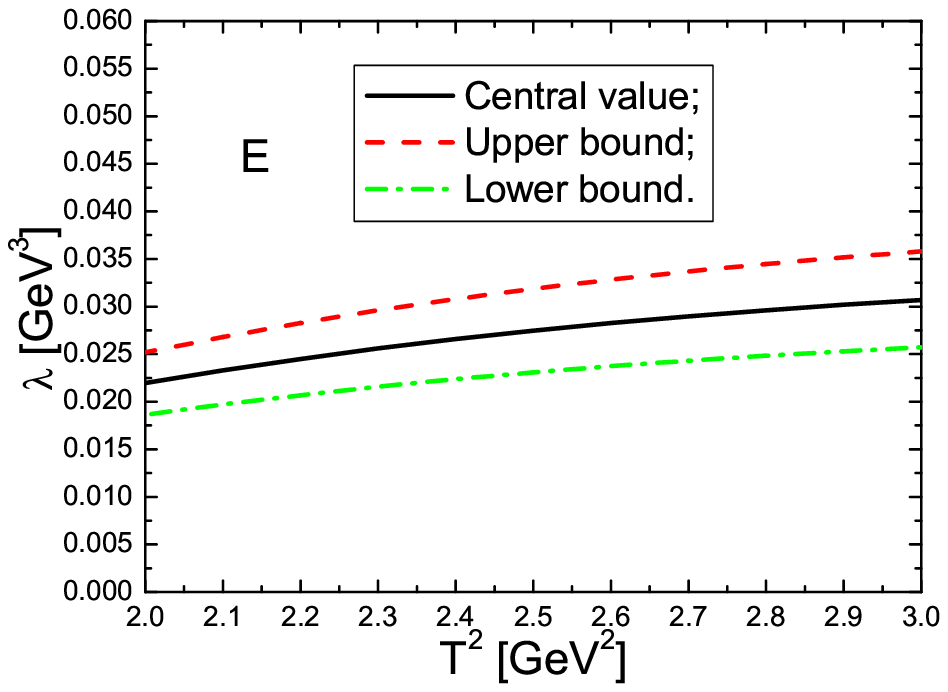}
 \includegraphics[totalheight=4cm,width=5cm]{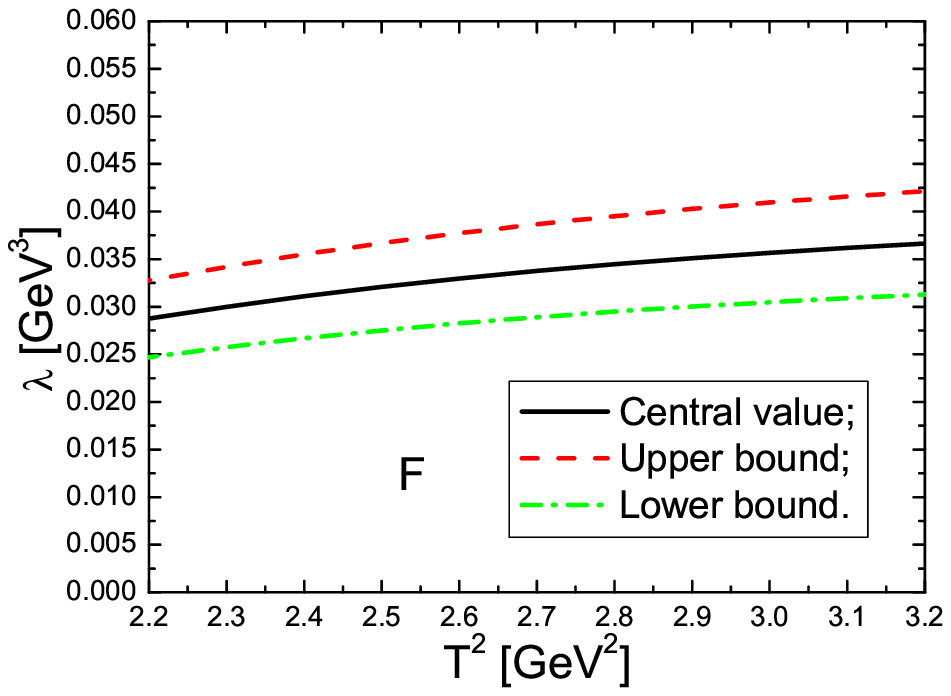}
\includegraphics[totalheight=4cm,width=5cm]{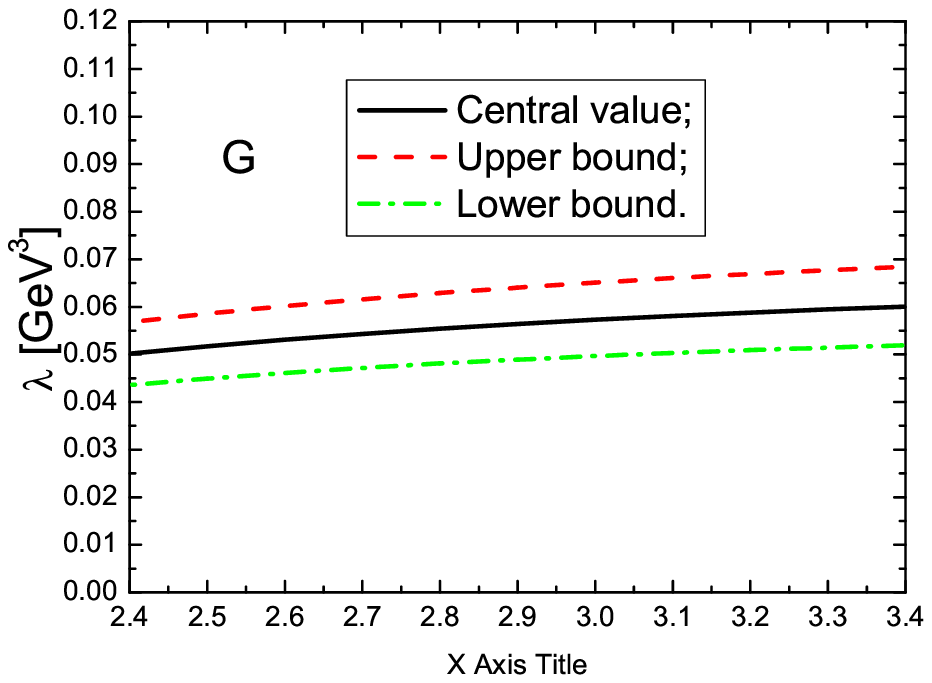}
\includegraphics[totalheight=4cm,width=5cm]{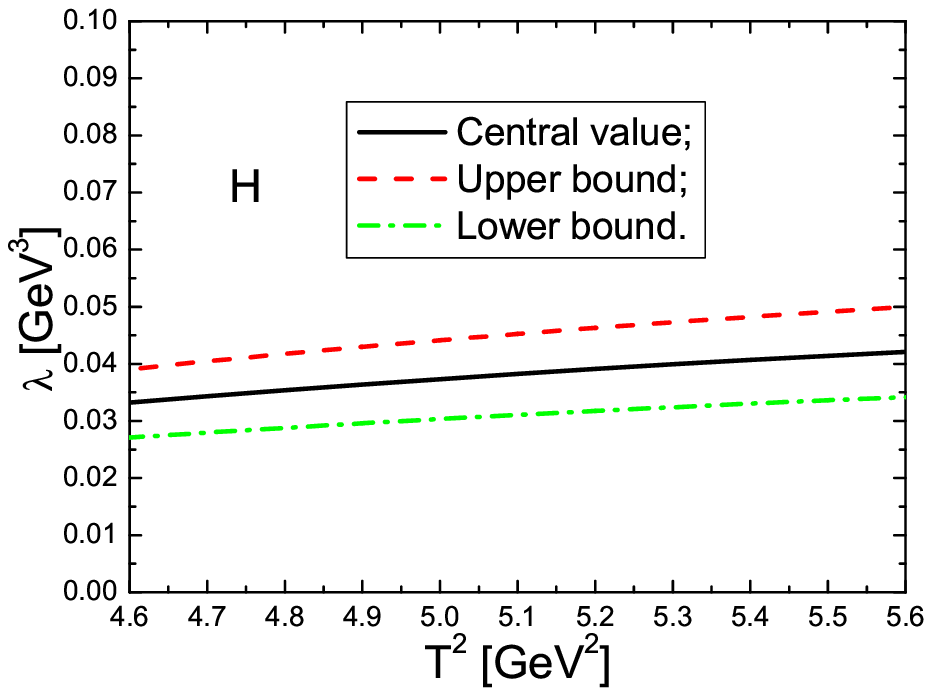}
\includegraphics[totalheight=4cm,width=5cm]{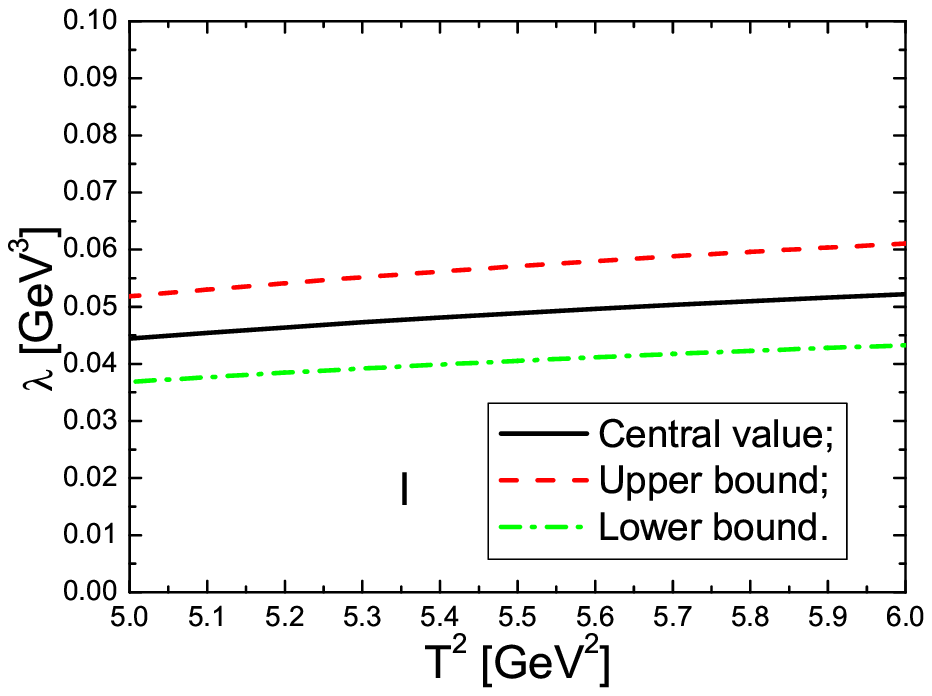}
 \includegraphics[totalheight=4cm,width=5cm]{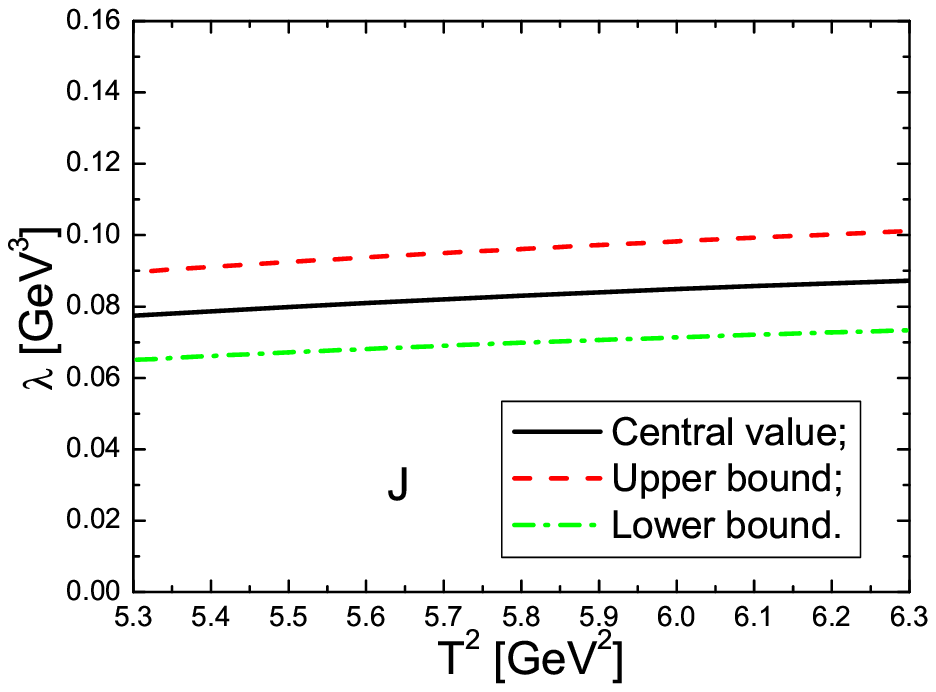}
        \caption{ The  pole residues of the heavy and doubly heavy baryon states, the
$A$, $B$, $C$, $D$, $E$, $F$, $G$, $H$, $I$ and $J$  correspond
      to the channels $\Xi^*_{cc}$, $\Omega^*_{cc}$, $\Xi^*_{bb}$,  $\Omega^*_{bb}$, $\Sigma_c^*$,
$\Xi_c^*$, $\Omega_c^*$, $\Sigma_b^*$, $\Xi_b^*$ and $\Omega_b^*$
respectively. }
\end{figure}

In this article, we take the simple pole $+$ continuum approximation
for the phenomenological spectral  densities. In fact, such a simple
approximation has shortcomings. In the case of the non-relativistic
harmonic-oscillator potential model,  the spectrum of the bound
states (the masses $E_n$ and the wave functions $\Psi_n(x)$) and the
exact correlation functions (and hence its operator product
expansion to any order) are known precisely. The non-relativistic
harmonic-oscillator potential $\frac{1}{2}m\omega^2\vec{r}^2$ is
highly non-perturbative, one suppose the full Green function
satisfies the Lippmann-Schwinger operator equation and may be solved
perturbatively. We can introduce the Borel parameter dependent
effective threshold parameter
$z_{eff}(T)=\omega\left[\bar{z}_0+\bar{z}_1 \sqrt{ \frac{\omega}{T}
}+\bar{z}_2 \frac{\omega}{T}+\cdots \right]$ and fit the
coefficients $\bar{z}_i$ to reproduce both the ground energy $E_0$
and the pole residue $R_0=\Psi_0^*(0)\Psi_0(0)$,  the
phenomenological spectrum density can be described by the
perturbative contributions well above the effective continuum
threshold  $z_{eff}(T)$, or reproduce the ground energy $E_0$ only
and take the pole residue $R$ as a calculated parameter, there
exists a solution for the effective continuum threshold $z_{eff}(T)$
which precisely reproduces the exact ground energy $E_0$
 for any value of the pole residue $R$ within the range $R=(0.7-1.15)R_0$ in the
limited fiducial Borel window,   the value of the pole residue $R$
 extracted from the sum rule is
determined to a great extent by the contribution of the hadron
continuum \cite{ChSh3}. There maybe systemic uncertainties  out of
control.

In the real QCD world, the hadronic spectral densities are not known
exactly. In the present case, the ground states in some channels
have not been observed yet. So we have no confidence to introduce
the Borel parameter dependent effective threshold parameter
$s_{eff}(T)=\bar{s}_0+\bar{s}_1 \frac{1}{T^2} +\bar{s}_2
\frac{1}{T^4}+\cdots $ and approximate the phenomenological spectral
densities  with the perturbative contributions above the effective
continuum  threshold $s_{eff}(T)$ accurately. Furthermore, the pole
residues (or the couplings of the interpolating  currents to the
ground state baryons) $\lambda_{\pm}$ are not experimentally
measurable quantities, and should be calculated by some theoretical
approaches, the true values are difficult to obtain,  which are
distinguished from the decay constants of the pseudoscalar mesons
and the vector mesons, the decay constants can be measured with
great precision in the leptonic decays (in some channels).

The  spectrum of the bound states in the non-relativistic
harmonic-oscillator potential model are of the Dirac $\delta$
function type, we can choose $z_{eff}<E_1$, while in the case of the
QCD, the situation is  rather complex, the effective continuum
thresholds $s_{eff}(T)$ maybe overlap with the first radial excited
states, which are usually broad. For example, in the pseudoscalar
channels, the widths of the  $\pi$, $\pi(1300)$, $\pi(1800)$,
$\cdots$ are $\sim 0\,\rm{GeV}$, $(0.2-0.6)\,\rm{GeV}$,
$0.208\pm0.012\,\rm{GeV}$, $\cdots$ respectively, while  the widths
of the $K$, $K(1460)$, $K(1830)$, $\cdots$ are $\sim 0 \,\rm{GeV}$,
$\sim (0.25-0.26)\,\rm{GeV}$, $\sim 0.25\,\rm{GeV}$, $\cdots$
respectively \cite{PDG}. In this article, we prefer (or have to
choose) the simple pole $+$ continuum approximation, and cannot
estimate the unknown systemic uncertainties of the QCD sum rules
before the spectral densities in both the QCD and phenomenological
sides are  known with great accuracy.

The properties of the charm and doubly charm baryon states would be
studied at the BESIII and $\rm{\bar{P}ANDA}$ \cite{BESIII,PANDA},
where the charm baryon states are copiously produced at the $e^+e^-$
and $p\bar{p}$ collisions. The LHCb is a dedicated $b$ and
$c$-physics precision experiment at the LHC (large hadron collider).
The LHC will be the world's most copious  source of the $b$ hadrons,
and  a complete spectrum of the $b$ hadrons will be available
through gluon fusion. In proton-proton collisions at
$\sqrt{s}=14\,\rm{TeV}$, the $b\bar{b}$ cross section is expected to
be $\sim 500\mu b$ producing $10^{12}$ $b\bar{b}$ pairs in a
standard  year of running at the LHCb operational luminosity of
$2\times10^{32} \rm{cm}^{-2} \rm{sec}^{-1}$ \cite{LHC}. The present
predictions for the masses of the heavy and doubly heavy baryon
states can be confronted with the experimental data  in the future
at the BESIII, $\rm{\bar{P}ANDA}$ and LHCb.

\section{Conclusion}
In this article, we study the  ${3\over 2}^+$ heavy and doubly heavy
baryon states $\Xi^*_{cc}$, $\Omega^*_{cc}$, $\Xi^*_{bb}$,
$\Omega^*_{bb}$, $\Sigma_c^*$, $\Xi_c^*$, $\Omega_c^*$,
$\Sigma_b^*$, $\Xi_b^*$ and $\Omega_b^*$   by subtracting the
contributions from the corresponding ${3\over 2}^-$ heavy and doubly
heavy baryon states with the QCD sum rules, and make reasonable
predictions for their masses.  The present predictions  can be
confronted with the experimental data in the future at the BESIII,
$\rm{\bar{P}ANDA}$ and LHCb, especially the LHCb. Once reasonable
values of the pole residues $\lambda_{+}$ are obtained, we can take
them as   basic input parameters and study the revelent hadronic
processes with the QCD sum rules.

\section*{Acknowledgements}
This  work is supported by National Natural Science Foundation,
Grant Number 10775051, and Program for New Century Excellent Talents
in University, Grant Number NCET-07-0282, and the Fundamental
Research Funds for the Central Universities.

\section*{Appendix}
The spectral densities of the heavy and doubly heavy baryon states
$\Omega^*_{QQ}$, $\Xi^*_{QQ}$, $\Omega^*_Q$, $\Xi^*_Q$ and
$\Sigma^*_Q$ at the level of quark-gluon degrees of freedom,

\begin{eqnarray}
\rho^A_{\Omega_{QQ}}(p_0)&=&\frac{3p_0}{16 \pi^4}
\int_{\alpha_{i}}^{\alpha_{f}}d\alpha \int_{\beta_{i}}^{1-\alpha}
d\beta\alpha\beta(1-\alpha-\beta)(p_0^2-\widetilde{m}^2_Q)(2p_0^2-\widetilde{m}^2_Q)
\nonumber\\
&&+\frac{3m_Q^2p_0}{16\pi^4}\int_{\alpha_{i}}^{\alpha_{f}}d\alpha
\int_{\beta_{i}}^{1-\alpha} d\beta
(1-\alpha-\beta)(p_0^2-\widetilde{m}^2_Q) \nonumber\\
&&-\frac{m_Q^2}{192\pi^2}
\langle\frac{\alpha_sGG}{\pi}\rangle\int_{\alpha_{i}}^{\alpha_{f}}d\alpha
\int_{\beta_{i}}^{1-\alpha} d\beta (1-\alpha-\beta)
\left[\frac{\alpha}{\beta^2}+\frac{\beta}{\alpha^2} \right]\left[1+\frac{p_0}{2T}\right]\delta(p_0-\widetilde{m}_Q)\nonumber\\
&&-\frac{m_Q^4}{384\pi^2p_0T}\langle\frac{\alpha_sGG}{\pi}\rangle\int_{\alpha_{i}}^{\alpha_{f}}d\alpha
\int_{\beta_{i}}^{1-\alpha} d\beta (1-\alpha-\beta)\left[\frac{1}{\alpha^3}+\frac{1}{\beta^3}\right]\delta(p_0-\widetilde{m}_Q)\nonumber\\
&&+\frac{m_Q^2}{64\pi^2}\langle\frac{\alpha_sGG}{\pi}\rangle\int_{\alpha_{i}}^{\alpha_{f}}d\alpha
\int_{\beta_{i}}^{1-\alpha} d\beta (1-\alpha-\beta)\left[\frac{1}{\alpha^2}+\frac{1}{\beta^2}\right]\delta(p_0-\widetilde{m}_Q)\nonumber\\
&&+\frac{m_s\langle\bar{s}{s}\rangle}{4\pi^2}\int_{\alpha_{i}}^{\alpha_{f}}d\alpha
\alpha(1-\alpha)\left[
p_0+\frac{p_0^2}{4}\delta(p_0-\widetilde{\widetilde{m}}_Q)\right] \nonumber\\
&&+\frac{m_sm_Q^2\langle\bar{s}{s}\rangle}{16\pi^2}\int_{\alpha_{i}}^{\alpha_{f}}d\alpha
\delta(p_0-\widetilde{\widetilde{m}}_Q)  \nonumber \\
&&-\frac{1}{48\pi^2}\langle\frac{\alpha_sGG}{\pi}\rangle\int_{\alpha_{i}}^{\alpha_{f}}d\alpha
\int_{\beta_{i}}^{1-\alpha} d\beta (1-\alpha-\beta)
\left[p_0+\frac{p_0^2}{8}\delta(p_0-\widetilde{m}_Q) \right]\, ,\\
\rho^B_{\Omega_{QQ}}(p_0)&=&\frac{3m_s}{32 \pi^4}
\int_{\alpha_{i}}^{\alpha_{f}}d\alpha \int_{\beta_{i}}^{1-\alpha}
d\beta\alpha\beta(p_0^2-\widetilde{m}^2_Q)(3p_0^2-2\widetilde{m}^2_Q)
\nonumber\\
&&+\frac{3m_sm_Q^2}{16\pi^4}\int_{\alpha_{i}}^{\alpha_{f}}d\alpha
\int_{\beta_{i}}^{1-\alpha} d\beta
(p_0^2-\widetilde{m}^2_Q) \nonumber\\
&&-\frac{m_sm_Q^2}{384\pi^2T}
\langle\frac{\alpha_sGG}{\pi}\rangle\int_{\alpha_{i}}^{\alpha_{f}}d\alpha
\int_{\beta_{i}}^{1-\alpha} d\beta
\left[\frac{\alpha}{\beta^2}+\frac{\beta}{\alpha^2} \right] \delta(p_0-\widetilde{m}_Q)\nonumber\\
&&-\frac{m_sm_Q^4}{384\pi^2p_0^2T}\langle\frac{\alpha_sGG}{\pi}\rangle\int_{\alpha_{i}}^{\alpha_{f}}d\alpha
\int_{\beta_{i}}^{1-\alpha} d\beta \left[\frac{1}{\alpha^3}+\frac{1}{\beta^3}\right]\delta(p_0-\widetilde{m}_Q)\nonumber\\
&&+\frac{m_sm_Q^2}{64\pi^2p_0}\langle\frac{\alpha_sGG}{\pi}\rangle\int_{\alpha_{i}}^{\alpha_{f}}d\alpha
\int_{\beta_{i}}^{1-\alpha} d\beta \left[\frac{1}{\alpha^2}+\frac{1}{\beta^2}\right]\delta(p_0-\widetilde{m}_Q)\nonumber\\
&&-\frac{\langle\bar{s}{s}\rangle}{4\pi^2}\int_{\alpha_{i}}^{\alpha_{f}}d\alpha
\alpha(1-\alpha)\left[ 2p_0^2-\widetilde{\widetilde{m}}_Q^2\right]
-\frac{m_Q^2\langle\bar{s}{s}\rangle}{4\pi^2}\int_{\alpha_{i}}^{\alpha_{f}}d\alpha
 \nonumber \\
&&-\frac{m_s}{64\pi^2}\langle\frac{\alpha_sGG}{\pi}\rangle\int_{\alpha_{i}}^{\alpha_{f}}d\alpha
\int_{\beta_{i}}^{1-\alpha} d\beta
\left[1+\frac{p_0}{6}\delta(p_0-\widetilde{m}_Q) \right] \, ,
\end{eqnarray}

\begin{eqnarray}
\rho^A_{\Xi_{QQ}}(p_0)&=&\frac{3p_0}{16 \pi^4}
\int_{\alpha_{i}}^{\alpha_{f}}d\alpha \int_{\beta_{i}}^{1-\alpha}
d\beta\alpha\beta(1-\alpha-\beta)(p_0^2-\widetilde{m}^2_Q)(2p_0^2-\widetilde{m}^2_Q)
\nonumber\\
&&+\frac{3m_Q^2p_0}{16\pi^4}\int_{\alpha_{i}}^{\alpha_{f}}d\alpha
\int_{\beta_{i}}^{1-\alpha} d\beta
(1-\alpha-\beta)(p_0^2-\widetilde{m}^2_Q) \nonumber\\
&&-\frac{m_Q^2}{192\pi^2}
\langle\frac{\alpha_sGG}{\pi}\rangle\int_{\alpha_{i}}^{\alpha_{f}}d\alpha
\int_{\beta_{i}}^{1-\alpha} d\beta (1-\alpha-\beta)
\left[\frac{\alpha}{\beta^2}+\frac{\beta}{\alpha^2} \right]\left[1+\frac{p_0}{2T}\right]\delta(p_0-\widetilde{m}_Q)\nonumber\\
&&-\frac{m_Q^4}{384\pi^2p_0T}\langle\frac{\alpha_sGG}{\pi}\rangle\int_{\alpha_{i}}^{\alpha_{f}}d\alpha
\int_{\beta_{i}}^{1-\alpha} d\beta (1-\alpha-\beta)\left[\frac{1}{\alpha^3}+\frac{1}{\beta^3}\right]\delta(p_0-\widetilde{m}_Q)\nonumber\\
&&+\frac{m_Q^2}{64\pi^2}\langle\frac{\alpha_sGG}{\pi}\rangle\int_{\alpha_{i}}^{\alpha_{f}}d\alpha
\int_{\beta_{i}}^{1-\alpha} d\beta (1-\alpha-\beta)\left[\frac{1}{\alpha^2}+\frac{1}{\beta^2}\right]\delta(p_0-\widetilde{m}_Q)\nonumber\\
&&-\frac{1}{48\pi^2}\langle\frac{\alpha_sGG}{\pi}\rangle\int_{\alpha_{i}}^{\alpha_{f}}d\alpha
\int_{\beta_{i}}^{1-\alpha} d\beta (1-\alpha-\beta)
\left[p_0+\frac{p_0^2}{8}\delta(p_0-\widetilde{m}_Q) \right]\, , \\
\rho^B_{\Xi_{QQ}}(p_0)&=&-\frac{\langle\bar{q}{q}\rangle}{4\pi^2}\int_{\alpha_{i}}^{\alpha_{f}}d\alpha
\alpha(1-\alpha)\left[ 2p_0^2-\widetilde{\widetilde{m}}_Q^2\right]
-\frac{m_Q^2\langle\bar{q}{q}\rangle}{4\pi^2}\int_{\alpha_{i}}^{\alpha_{f}}d\alpha
  \, ,
\end{eqnarray}

\begin{eqnarray}
\rho^A_{\Omega_Q}(p_0)&=&\frac{p_0}{64\pi^4}\int_{t_i}^1dt
t(2+t)(1-t)^2(p_0^2-\widetilde{m}_Q^2)^2-\frac{p_0m_s\langle\bar{s}s\rangle}{4\pi^2}\int_{t_i}^1
dt t(2-t)\nonumber\\
&&+\frac{m_s\langle\bar{s}g_s\sigma Gs\rangle}{48\pi^2}\int_0^1dt
t\delta (p_0-\widetilde{m}_Q) +\frac{m_s\langle\bar{s}g_s\sigma
Gs\rangle}{24\pi^2}\delta
(p_0-m_Q)+\frac{\langle\bar{s}s\rangle^2}{6}\delta(p_0-m_Q)\nonumber \\
&&-\frac{p_0}{192\pi^2}\langle \frac{\alpha_sGG}{\pi}\rangle
\int_{t_i}^1 dt t(2-t) +\frac{m_Q^2}{1152\pi^2}\langle
\frac{\alpha_sGG}{\pi}\rangle \int_{t_i}^1 dt
\frac{(1-t)^3}{t^2}\delta (p_0-\widetilde{m}_Q)
\nonumber \\
&& -\frac{m_Q^2}{384\pi^2}\langle \frac{\alpha_sGG}{\pi}\rangle
\int_0^1 dt\frac{(1-t)^2}{t^2}\delta (p_0-\widetilde{m}_Q) \, ,
\\
\rho^B_{\Omega_Q}(p_0)&=&\frac{m_Q}{64\pi^4}\int_{t_i}^1dt
(2+t)(1-t)^2(p_0^2-\widetilde{m}_Q^2)^2-\frac{m_sm_Q\langle\bar{s}s\rangle}{4\pi^2}\int_{t_i}^1
dt (2-t)
\nonumber\\
&&+\frac{m_sm_Q\langle\bar{s}g_s\sigma
Gs\rangle}{48\pi^2p_0}\int_0^1dt \delta
(p_0-\widetilde{m}_Q)+\frac{m_s \langle\bar{s}g_s\sigma Gs\rangle
}{24\pi^2}\delta
(p_0-m_Q) +\frac{\langle\bar{s}s\rangle^2}{6}\delta(p_0-m_Q)\nonumber\\
&&-\frac{m_Q}{192\pi^2}\langle \frac{\alpha_sGG}{\pi}\rangle
\int_{t_i}^1 dt (2-t)+\frac{m_Q}{576\pi^2}\langle
\frac{\alpha_sGG}{\pi}\rangle
\int_{t_i}^1 dt \frac{(1-t)^3(3t+4)}{t^2}\nonumber \\
&& -\frac{m_Q}{1152\pi^2}\langle \frac{\alpha_sGG}{\pi}\rangle
\int_0^1 dt\frac{t^3-3t+2}{t}\widetilde{m}_Q \delta
(p_0-\widetilde{m}_Q) \, ,
\end{eqnarray}

\begin{eqnarray}
\rho^A_{\Xi_Q}(p_0)&=&\frac{p_0}{128\pi^4}\int_{t_i}^1dt
t(2+t)(1-t)^2(p_0^2-\widetilde{m}_Q^2)^2+\frac{p_0m_s\langle\bar{s}s\rangle}{16\pi^2}\int_{t_i}^1
dt t^2\nonumber\\
&&-\frac{m_s\langle\bar{q}q\rangle}{8\pi^2}\int_{t_i}^1 dt t
+\frac{m_s\langle\bar{s}g_s\sigma Gs\rangle}{192\pi^2}\int_0^1dt t
\delta (p_0-\widetilde{m}_Q ) \nonumber\\
&&+\frac{m_s \left[3\langle\bar{q}g_s\sigma
Gq\rangle-\langle\bar{s}g_s\sigma Gs\rangle \right]}{192\pi^2}\delta
(p_0-m_Q)+\frac{\langle\bar{q}q\rangle\langle\bar{s}s\rangle}{12}\delta(p_0-m_Q)\nonumber \\
&& -\frac{p_0}{384\pi^2}\langle \frac{\alpha_sGG}{\pi}\rangle
\int_{t_i}^1 dt t(2-t) \nonumber \\
&& -\frac{m_Q^2}{2304\pi^2}\langle \frac{\alpha_sGG}{\pi}\rangle
\int_{t_i}^1 dt \frac{t^3-3t+2}{t^2}\delta (p_0-\widetilde{m}_Q) \,
, \\
\rho^B_{\Xi_Q}(p_0)&=&\frac{m_Q}{128\pi^4}\int_{t_i}^1dt
(2+t)(1-t)^2(p_0^2-\widetilde{m}_Q^2)^2+\frac{m_sm_Q\langle\bar{s}s\rangle}{16\pi^2}\int_{t_i}^1
dt t- \frac{m_sm_Q\langle\bar{q}q\rangle}{8\pi^2}\int_{t_i}^1 dt
\nonumber\\
&&+\frac{m_sm_Q\langle\bar{s}g_s\sigma
Gs\rangle}{192\pi^2p_0}\int_0^1dt t \delta (p_0-\widetilde{m}_Q
)+\frac{m_s\left[3\langle\bar{q}g_s\sigma
Gq\rangle-\langle\bar{s}g_s\sigma Gs\rangle \right]}{192\pi^2}\delta
(p_0-m_Q)\nonumber\\
 && +\frac{\langle\bar{q}q\rangle\langle\bar{s}s\rangle}{12}\delta(p_0-m_Q)-\frac{m_Q}{384\pi^2}\langle \frac{\alpha_sGG}{\pi}\rangle
\int_{t_i}^1 dt (2-t)\nonumber\\
&&+\frac{m_Q}{1152\pi^2}\langle \frac{\alpha_sGG}{\pi}\rangle
\int_{t_i}^1 dt \frac{(1-t)^3(3t+4)}{t^2}\nonumber \\
&& -\frac{m_Q}{2304\pi^2}\langle \frac{\alpha_sGG}{\pi}\rangle
\int_0^1 dt\frac{t^3-3t+2}{t}\widetilde{m}_Q \delta
(p_0-\widetilde{m}_Q) \, ,
\end{eqnarray}

\begin{eqnarray}
\rho^A_{\Sigma_Q}(p_0)&=&\frac{p_0}{128\pi^4}\int_{t_i}^1dt
t(2+t)(1-t)^2(p_0^2-\widetilde{m}_Q^2)^2+\frac{\langle\bar{q}q\rangle^2}{12}\delta(p_0-m_Q)\nonumber \\
&& -\frac{p_0}{384\pi^2}\langle \frac{\alpha_sGG}{\pi}\rangle
\int_{t_i}^1 dt t(2-t) \nonumber \\
&& -\frac{m_Q^2}{2304\pi^2}\langle \frac{\alpha_sGG}{\pi}\rangle
\int_{t_i}^1 dt \frac{t^3-3t+2}{t^2}\delta (p_0-\widetilde{m}_Q) \,
, \\
\rho^B_{\Sigma_Q}(p_0)&=&\frac{m_Q}{128\pi^4}\int_{t_i}^1dt
(2+t)(1-t)^2(p_0^2-\widetilde{m}_Q^2)^2+\frac{\langle\bar{q}q\rangle^2}{12}\delta(p_0-m_Q)
\nonumber\\
 && -\frac{m_Q}{384\pi^2}\langle \frac{\alpha_sGG}{\pi}\rangle
\int_{t_i}^1 dt t(2-t)+\frac{m_Q}{1152\pi^2}\langle
\frac{\alpha_sGG}{\pi}\rangle
\int_{t_i}^1 dt \frac{(1-t)^3(3t+4)}{t^2}\nonumber \\
&& -\frac{m_Q}{2304\pi^2}\langle \frac{\alpha_sGG}{\pi}\rangle
\int_0^1 dt\frac{t^3-3t+2}{t}\widetilde{m}_Q \delta
(p_0-\widetilde{m}_Q) \, ,
\end{eqnarray}
where $\alpha_{f}=\frac{1+\sqrt{1-4m_Q^2/p_0^2}}{2}$,
$\alpha_{i}=\frac{1-\sqrt{1-4m_Q^2/p_0^2}}{2}$,
$\beta_{i}=\frac{\alpha m_Q^2}{\alpha p_0^2 -m_Q^2}$,
$\widetilde{m}_Q^2=\frac{(\alpha+\beta)m_Q^2}{\alpha\beta}$,
$\widetilde{\widetilde{m}}_Q^2=\frac{m_Q^2}{\alpha(1-\alpha)}$
   in the channels $\Omega^*_{QQ}$ and $\Xi^*_{QQ}$; and $\widetilde{m}_Q^2=\frac{m_Q^2}{t}$,
$t_i=\frac{m_Q^2}{p_0^2}$     in the channels $\Omega^*_Q$,
$\Xi^*_Q$ and $\Sigma_Q^*$.

\end{document}